\documentstyle[epsfig,aps,prb]{revtex}
\newcommand\beq{\begin{equation}}
\newcommand\eeq{\end{equation}}
\newcommand\bea{\begin{eqnarray}}
\newcommand\eea{\end{eqnarray}}
\newcommand\jp{J^\prime }

\begin{document}

\begin{center}
{\Large Magnetization properties of some quantum spin ladders}
\end{center}

\vskip .5 true cm
\centerline{\bf Kunj Tandon$^1$, Siddhartha Lal$^2$, Swapan K. Pati$^1$, S. 
Ramasesha$^1$}
\centerline{\bf and Diptiman Sen$^3$} 
\vskip .5 true cm

\centerline{\it $^1$ Solid State and Structural Chemistry Unit} 
\centerline{\it $^2$ Department of Physics}  
\centerline{\it $^3$ Centre for Theoretical Studies}  
\centerline{\it Indian Institute of Science, Bangalore 560012, India} 
\vskip .5 true cm

\begin{abstract}

The experimental realization of various spin ladder systems has prompted their 
detailed theoretical investigations. Here we study the evolution of 
ground-state magnetization with an external magnetic field for two different 
antiferromagnetic systems: a three-legged spin-$1/2$ ladder, and a two-legged 
spin-$1/2$ ladder with an additional diagonal interaction. The finite system 
density-matrix renormalization-group method is employed for numerical studies 
of the three-chain system, and an effective low-energy Hamiltonian is used in 
the limit of strong interchain coupling to study the two- and three-chain 
systems. The three-chain system has a magnetization plateau at one-third of 
the saturation magnetization. The two-chain system has a plateau at zero 
magnetization due to a gap above the singlet ground state. It also has a 
plateau at half of the saturation magnetization for a certain range of values 
of the couplings. We study the regions of transitions between plateaus 
numerically and analytically, and find that they are described, at first order 
in a strong-coupling expansion, by an $XXZ$ spin-$1/2$ chain in a magnetic 
field; the second-order terms give corrections to the $XXZ$ model. We also
study numerically some low-temperature properties of the three-chain 
system, such as the magnetization, magnetic susceptibility and specific heat.

\end{abstract}
\vskip .5 true cm

~~~~~~ PACS number: ~75.10.Jm, ~75.60.Ej

\newpage

\section{Introduction}

One-dimensional and quasi-one-dimensional quantum spin systems have been
studied extensively in recent years for several reasons. Many such systems
have been realized experimentally, and a variety of theoretical techniques, 
both analytical and numerical, are available to study the relevant
models. Due to large quantum fluctuations in low dimensions, 
such systems often have unusual properties such as a gap between a singlet 
ground state and the excited nonsinglet states; this leads to a 
magnetic susceptibility which vanishes exponentially at low temperatures.
Perhaps the most famous example of this is the Haldane gap which was
predicted theoretically in integer spin Heisenberg antiferromagnetic
chains \cite{hald1}, and then observed experimentally in a spin-$1$ system
$Ni(C_2 H_8 N_2 )_2 NO_2 (Cl O_4 )$ \cite{buye}. Other examples include the
spin ladder systems in which a small number of one-dimensional spin-$1/2$
chains interact amongst each other \cite{dago}. It has been observed that 
if the number of chains is even, i.e., if each rung of the ladder (which is
the unit cell for the system) contains an even number
of spin-$1/2$ sites, then the system effectively behaves like an integer spin
chain with a gap in the low-energy spectrum. Some two-chain ladders which 
show a gap are $(VO)_2 P_2 O_7$ \cite{eccl}, $Sr Cu_2 O_3$ \cite{azum}
and $Cu_2 (C_5 H_{12} N_2 )_2 Cl_4$ \cite{chab1}. Conversely, a three-chain
ladder which effectively behaves like a half-odd-integer spin chain and 
does {\it not} exhibit a gap is $Sr_2 Cu_3 O_5$ \cite{azum}. A related
observation is that the quasi-one-dimensional system $CuGeO_3$ spontaneously
dimerizes below a spin-Peierls transition temperature \cite{hase}; then 
the unit cell contains two spin-$1/2$ sites and the system is gapped.

The results for gaps quoted above are all in the absence of an external 
magnetic field. The situation becomes more interesting in the 
presence of a magnetic field \cite{chab2}. Then it is possible for an 
integer spin chain to be gapless and a half-odd-integer spin chain to show 
a gap above the ground state for appropriate values of the field 
\cite{oshi,cabr1,cabr2,hida,kura,tots1,tots2,tone,saka1,saka2,chit1}. This has
been demonstrated in several models using a variety of methods such as exact
diagonalization of small systems, bosonization and conformal field theory
\cite{schu,affl}, and perturbation theory \cite{reig}. In particular, it has 
been shown that the magnetization of the system can exhibit plateaus at 
certain nonzero values for some finite ranges of the magnetic field. Further, 
for a Hamiltonian which is invariant under translation by one unit cell, the 
value of the magnetization per unit cell is quantized to be a rational number 
at each plateau. 

The necessary (but not sufficient) condition for the magnetization
quantization is given as follows \cite{oshi}. Let us assume that the magnetic 
field points along the $\hat z$ axis, the total Hamiltonian $H$ is invariant 
under spin rotations about that axis, and the maximum possible spin in each 
unit cell of the Hamiltonian is given by $S$. Consider a state $\psi$ such
that the expectation value of $S_z$ per unit cell is equal to $m_s$ in that 
state, and $\psi$ has a period $n$, i.e., it is invariant only under 
translation by a number of unit cells equal to $n$ or a multiple of $n$. (It 
is clear that if $n \ge 2$, then there must be $n$ such states with the same
energy, since $H$ is invariant under a translation by one unit cell). Then
the quantization condition says that a magnetic plateau is possible at
the state $\psi$, i.e., there is a range of values of the external field for
which $\psi$ is the ground state and is separated by a finite gap from states
with slightly higher or lower values of total $S_z$, only if 
\beq
n ~(~ S ~-~ m_s ~) ~=~ {\rm an ~~ integer} .
\label{quant}
\eeq
This condition is very useful because it enables us to restrict our attention
to some particular values of $m_s$ and $n$ when searching for possible plateaus 
in a given model. Note that the saturated state in which all spins point 
along the magnetic field trivially satisfies (\ref{quant}) since it has
$m_s =S$ (or $-S$) and $n=1$.

In this paper, we will study the magnetization
as a function of the applied field for a two- and three-chain ladder. We will
do so both numerically, using the density-matrix renormalization group method 
(DMRG) \cite{whit,kawa}, and perturbatively, using a low-energy effective 
Hamiltonian (LEH) \cite{tots2,mila}. Our analysis will 
extend the currently known results in many ways. We have used
DMRG to study two-spin correlation functions in the ground state, and some 
finite-temperature thermodynamic properties such as magnetic susceptibility 
and specific heat. Further, our LEH goes up to the second order in a 
strong-coupling expansion. Whenever possible, we will 
use the analytical results from the LEH to understand the numerical results. 
The first-order LEH will turn out to be the well-studied $XXZ$ spin-$1/2$ 
chain in a longitudinal magnetic field \cite{cabr2,hald2}, and it will 
usually prove to be sufficient 
for a qualitative understanding of the results. However, we will find it 
necessary to invoke the second-order results (which give corrections to
the $XXZ$ model) for a more accurate comparison with the numerics.

The paper is organized as follows. In Sec. II, we will present
all the numerical results we have obtained for the three-chain ladder
using DMRG. We will see that there is a finite energy gap and exponentially
decaying spin correlations at each plateau, while there is no gap and
the two-spin correlations decay as powers 
in between two plateaus. We will also study how the plateaus gradually
disappear and how the susceptibility and specific heat evolve as we 
increase the temperature. In Sec. III, we will derive the LEH for the same
model and show how it can be
used to understand some of the numerical results in Sec. II. 
We will also derive the LEH for a two-chain ladder which can be thought 
of as a dimerized and frustrated spin-$1/2$ chain \cite{chit2}, and we
will use it to understand magnetization plateaus in the ground state. We will
see that for certain values of the dimerization and frustration, the 
ground state can spontaneously break translation invariance leading to 
an additional plateau at an intermediate value of the magnetization. In
Sec. IV, we will summarize our results and point out some directions for
future studies.

\section{Density-matrix renormalization-group study of the three-chain ladder}

We have numerically studied a three-chain spin-$1/2$ ladder governed by the
Hamiltonian 
\beq
H ~=~ \jp ~\sum_a ~\sum_n ~ {\bf S}_{a,n} \cdot {\bf S}_{a+1,n} ~+~ J ~
\sum_{a=1}^3 ~\sum_n ~{\bf S}_{a,n} \cdot {\bf S}_{a,n+1} ~
-~ h ~\sum_{a=1}^3 ~\sum_n ~S_{a,n}^z ~,
\label{ham1}
\eeq
where $a$ denotes the chain index, $n$ denotes the rung index, $h$ denotes 
the magnetic field (we have absorbed the gyromagnetic ratio 
$g$ and the Bohr magneton $\mu_B$ in the definition of $h$), and $J, \jp > 0$. 
For convenience, we choose $h \ge 0$ since the region $h < 0$ can be deduced 
from it by reflection about the zero field. It is convenient to scale out 
the parameter $J$, and quote all results in terms of the two dimensionless 
quantities $\jp /J$ and $h/J$. If the length of each chain is $L$, 
the total number of sites is $N = 3L$. Since the total $S^z$ is a good quantum 
number, it is more convenient to do the numerical computations {\it without}
including the magnetic-field term in (\ref{ham1}), and then to add the effect
of the field at the end of the computation. The label $n+1$ (or $a+1$) is
appropriately interpreted for periodic boundary conditions along the chain
(or rung).

For the ground state properties, we have only considered an open boundary 
condition (OBC) in the rung direction, namely, the summation over $a$ in the 
first term of (\ref{ham1}) runs over $1,2$. However, for low-temperature 
properties, we have studied both OBC, as well as a periodic boundary condition 
(PBC) in the rung direction in which we sum over $a=1,2,3$ in the first term. 
(Only the OBC is realized along the rungs in the experimental systems studied 
so far. However, PBC along the rungs is interesting for theoretical reasons as 
we will see below).

For small systems, we have performed exact diagonalization with periodic 
boundary conditions in the chain direction. For larger systems, we have done 
DMRG calculations (using the finite system algorithm \cite{whit})
with open boundary conditions in the chain
direction. For exact diagonalization, we have gone up to $24$ sites, i.e., a 
chain length of $8$. With DMRG, we have gone up to $120$ sites (chain length of 
$40$) after checking that the DMRG and exact results match for $24$ sites. 
The number of dominant density matrix eigenstates, corresponding to
the $m$ largest eigenvalues of the density matrix, that we retained 
at each DMRG iteration was $m =80$. In fact, we varied the value of $m$ 
from $60$ to $100$, and found that $m =80$ gives satisfactory results in 
terms of agreement with exact diagonalization for small systems and good 
numerical convergence for large systems. 
For inputting the values of the couplings into the numerical 
programmes, it is more convenient to think of the system as a single chain 
(rather than as three chains) with the Hamiltonian
\beq
H ~=~ \frac{2}{3} ~\jp ~\sum_i ~[~ 1 ~-~ \cos ~(~\frac{2\pi i}{3}~) ~]~ {\bf
S}_i \cdot {\bf S}_{i+1} ~+~ J ~\sum_i ~{\bf S}_i \cdot {\bf S}_{i+3} ~.
\eeq
The system is grown by adding two new sites at each iteration.
Note that our method of construction ensures that we obtain the 
three-chain ladder structure after every third iteration when the total 
number of sites becomes a multiple of $6$. At various system sizes, starting
from $48$ sites and going up to $120$ sites in multiples of $6$ sites, we
computed the energies after doing three finite system iterations; we
found that the energy converges very well after three iterations. The energy
data is used in Figs. \ref{quadfit} and \ref{platgap} below. After reaching 
$120$ sites, we computed the spin densities and correlations after doing 
three finite system iterations. This data is used in Figs. \ref{spnden1} - 
\ref{spnden2}.

All our numerical results quoted below are for $J / \jp = 1/3$.  
We chose this particular value of the ratio for two reasons; 
there is a particularly broad magnetization plateau at $m_s =1/2$ which can be
easily found numerically, and that value of the 
ratio is sufficiently deep inside the strong-coupling regime that
the second-order perturbation expansion of Sec. II gives 
results which compare very well with the numerics.

We now describe the various ground-state properties we have found with OBC 
along the rungs. We looked for a magnetization plateau as follows. Motivated 
by the conditions in (\ref{quant}), we looked for a plateau at $m_s =1/2$ which 
would correspond to $n=1$ in that equation, since $S=3/2$. We also looked for 
plateaus at $m_s =0$ and $m_s =1$, each of which would correspond to $n=2$, 
i.e., a doubly degenerate state which has a period of two rungs. For a system 
with $N$ sites, a given value of magnetization per rung, $m_s$, corresponds 
to a sector with total $S^z$ equal to $M=m_s N/3$. Using the infinite
system algorithm, we found the lowest energies $E_0 (S^z,N)$ in the three 
sectors $S^z = M+1, M$ and $M-1$. Then we examined the 
three plots of $E_0 /NJ$ versus $1/N$ and extrapolated the
results up to the thermodynamic limit $N \rightarrow \infty$. We fitted these
plots with the formula $E_0 /NJ = e_i + a_i /N + b_i/N^2$, where the label 
$i=1,2,3$ denotes the $S^z$ sectors $M+1, M$ and $M-1$. (We found that a 
quadratic fit in $1/N$ matches the data much better than just a linear fit).
In the thermodynamic limit, the values of the three intercepts $e_i$ should 
match since those are just the energy per site for the three states whose 
$S_z$'s differ by only $1$. However, the three slopes $a_i$ are not equal 
in general. We now show that there is a magnetization plateau if $a_1 + a_3 
- 2 a_2$ has a nonzero value. Since the three energies $E_0$ are computed 
without including the magnetic field term, the upper critical field $h_{c+}$ 
where the states with $S_z =M+1$ and $M$ become degenerate is given by
\beq
h_{c+} (N) ~=~ E_0 (M+1, N) ~-~ E_0 (M, N) ~.
\label{hplus}
\eeq 
Similarly, the lower critical field $h_{c-}$ where the states with $S_z =M$ 
and $M-1$ become degenerate is given by
\beq
h_{c-} (N) ~=~ E_0 (M, N) ~-~ E_0 (M-1, N) ~.
\label{hminus}
\eeq 
\noindent
We therefore have a finite interval $\Delta h (N) = h_{c+} (N) - h_{c-} (N)$ 
in which the lowest energy state with $S^z = M$ is the ground state of the 
system with $N$ sites in the presence of a field $h$. If this interval 
has a nonzero limit as $N \rightarrow \infty$, we have a magnetization 
plateau. Thus, in the thermodynamic limit, the plateau width
$\Delta h /J$ is equal to $a_1 + a_3 - 2 a_2$. 

We will now quote our numerical results for $J / \jp =1/3$. For a rung 
magnetization of $m_s =1/2$, i.e., $M=N/6$, we found the three slopes
$a_i$ to be equal to $3.77, -0.02$ and $-1.93$; see Fig.
\ref{quadfit}. This gives the upper and lower critical fields to be
\bea
\frac{h_{c+}}{J}  ~=~ a_1 ~-~ a_2 ~&=&~ 3.79 ~, ~~~~~~~~ \frac{h_{c-}}{J} ~=~ 
a_2 ~-~ a_3 ~=~ 1.91 ~, \nonumber \\
\frac{\Delta h}{J} ~&=&~ \frac{h_{c+} ~-~ h_{c-}}{J} ~=~ 1.88 ~.
\label{plat1}
\eea
This is a sizeable plateau width, and it agrees with the exact diagonalization
results \cite{cabr1} and with the second-order perturbation expansion which 
will be discussed in the next section. For a rung magnetization of $m_s =1$,
we found the $a_i$ to be equal to $4.97, -0.24$ and $-5.43$.
Thus the upper and lower critical fields are 
\beq
\frac{h_{c+}}{J}  ~=~ 5.21 ~, ~~~~~~~~ \frac{h_{c-}}{J} ~=~ 5.19 ~, ~~~~~~~~
\frac{\Delta h}{J} ~=~ 0.02 ~.
\label{plat2}
\eeq
Finally, for a rung magnetization of $m_s =0$, we need the energies of
states with $M=0$ and $M= \pm 1$. Since the last two states must have the
same energy, we have $a_1 = a_3$ and it is sufficient to plot only $E_0 (0,N)$
and $E_0 (1,N)$ versus $1/N$. We found $a_1$ and $a_2$ to be equal to $0.39$
and $0.34$. This gives the upper and lower fields to be
\beq
\frac{h_{c+}}{J}  ~=~ 0.05 ~, ~~~~~~~~ \frac{h_{c-}}{J}  ~=~ -0.05 ~, ~~~~~~~~
\frac{\Delta h}{J} ~=~ 0.10 ~.
\label{plat3}
\eeq
The plateau widths given in (\ref{plat2}) and (\ref{plat3}) are rather small. 
In Fig. \ref{platgap}, we indicate the plateau widths $\Delta h (N) /J$ 
as a function of $1/N$ for $m_s =1/2$, $0$ and $1$. We will see that the LEH 
in the next section actually predicts that there should be no plateaus 
at $m_s =0$ and $1$.

Next, we computed the various spin correlations for the $120$-site system. We 
studied the spin densities $\langle S_{a,n}^z \rangle$ where the chain index 
$a=1,2,3$ and $n$ is the rung index. [Due to the rotation invariance about 
the $\hat z$ axis, the other two spin densities $\langle S_{a,n}^{\pm} 
\rangle$ must vanish]. For the plateau at $m_s =1/2$, we found that
\beq
\langle S_{1,n}^z \rangle ~=~ \langle S_{3,n}^z \rangle ~=~ 0.27 ~, ~~~~~~~~
\langle S_{2,n}^z \rangle ~=~ -0.04 ~, 
\eeq
for values of $n$ in the middle of the system. The spin densities are shown
in Fig. \ref{spnden1}. 

We also examined several 
two-spin correlations which can be denoted by $\langle S_{a,l}^z S_{b,n}^z 
\rangle$ and $\langle S_{a,l}^+ S_{b,n}^- \rangle$. For the $zz$
correlations, it is convenient to subtract the product of the two separate
spin densities; the subtracted $zz$ correlations then go to zero for large 
rung separations $\vert l - n \vert$, just like the $+-$ correlations. At 
$m_s =1/2$, we found that all these correlations decay very rapidly to zero as 
$\vert l - n \vert$ grows. In fact, the fall offs were so fast that we were 
unable to compute sensible correlation lengths. All the correlation lengths 
are of the order of one or two rungs as can be seen in Figs. \ref{corr22pm} 
and \ref{corr11zz}.

On the other hand,
for the state at $m_s =0$, we found that all the two-spin correlations
decay quite slowly. The decays are consistent with power law fall offs of the
form $A (-1)^{\vert l-n \vert} / \vert l-n \vert^{\eta}$. It is
difficult to find $\eta$ very accurately since the maximum value of $\vert 
l - n \vert$ is only $20$; this is because we fixed one site to be in the 
middle of the chain (to minimize edge effects), and the maximum chain length 
is $40$ for our DMRG calculations. For $m_s =0$, the exponent $\eta$ for 
all the correlations was found to be around $1$. There was no difference in 
the behaviors of the $zz$ and $+-$ correlations since this was an isotropic 
system; $m_s =0$ is the ground state if the magnetic field is zero.

For the state at $m_s =1$ (which is the ground state only for a substantial
value of the magnetic field), we found that the $+-$ correlations again
decay quite slowly consistent with a power law. The exponents $\eta$ for 
the different $+-$ correlations varied from $0.61$ to $0.70$ with an 
average value of $0.66$; see Fig. \ref{corr12pm} for an example. However, the
$zz$ correlations actually increased, rather than decreased, with increasing
separation $\vert l - n \vert$; see Fig. \ref{corr22zz}. We found that this 
is because of large edge effects. Since the magnetic field is particularly 
strong for the state with $m_s =1$, and sites at the ends have fewer neighbors 
coupled antiferromagnetically to them, they respond more strongly to the 
magnetic field than sites near the center of the system. This can be seen from 
Fig. \ref{spnden2} where the spin density $S_{a,n}^z$ shows a sharp increase 
towards the end of the chain (the rung index $n$ is equal to $1$ at the end).

We now summarize the properties of the three states studied with OBC
along the rungs. The state with $m_s =1/2$ is characterized by a large gap to 
excited states and extremely short correlation lengths for spin correlations. 
The states at $m_s =0$ and $m_s =1$ appear to have no gaps to excited states 
(within our numerical accuracy), and have slow fall offs of correlation 
functions consistent with power laws.

We now describe some low-temperature thermodynamic properties of the
three-chain system obtained using DMRG. Although DMRG is normally expected to 
be most accurate for targeting the lowest states in different $S^z$ sectors,
earlier studies of mixed spin chains have shown that DMRG is quite reliable
for computing low-temperature properties also \cite{pati}. There are two 
reasons for this; the low-lying excited states generally have a large 
projection onto the space of DMRG states which contains the ground state, and
the low-lying excitations in one sector are usually the lowest states in nearby
$S^z$ sectors. 

We first checked that for systems with $12$ sites, the results obtained 
using DMRG agree well with those obtained by exact diagonalization. We then 
used DMRG to study the magnetization, susceptibility and specific heat of
$36$-site systems with both OBC and PBC along the rungs. We first compute 
the partition function $Z = \sum_i ~{\rm exp}~ [-\beta (E_i - h (S^z)_i )]~$,
where the sum is over all the states $i$ in all the $S^z$ sectors, and
$\beta = 1/k_B T$ where $k_B$ is the Boltzmann constant. Then the 
magnetization is given by
\beq
\langle M \rangle ~=~ \frac{1}{Z} ~\sum_i ~(S^z)_i ~e^{-~\beta ~[~E_i ~-~ 
h ~( S^z)_i ~]} ~.
\eeq
The magnetic susceptibility is related to the fluctuation in magnetization,
\beq
\chi ~=~ \beta ~[~ \langle M^2 \rangle ~-~ \langle M \rangle^2 ~]~,
\eeq
and the specific heat is related to the fluctuation in energy,
\beq
\frac{C_V}{k_B} ~=~ \beta^2 ~[~ \langle E^2 \rangle ~-~ \langle E 
\rangle^2 ~]~.
\eeq

The plots of magnetization versus magnetic field for various temperatures
are shown in Figs. \ref{mo36} and \ref{mp36} for OBC and PBC, respectively,
along the rungs. 
The temperature $T$ is measured in units of $J/k_B$. We see that the 
plateau at $m_s =1/2$ disappears quite rapidly as we increase the temperature.
With OBC along rungs, the plateau has almost disappeared at $T=0.4$ which 
is substantially lower than the width $\Delta h/J = 1.88$. The plots of 
susceptibility in Fig. \ref{chio36} for OBC also show 
no surprises. The susceptibility is (exponentially) small at low temperatures 
in the region of the plateau because the magnetic excitations there are 
separated from the ground state by a gap. 

However, the specific heats shown in Figs. \ref{cvo36} and \ref{cvp36}
demonstrate an interesting difference between OBC and PBC along the rungs. 
While it is very small at low temperatures for OBC, it is not small for PBC; 
further, it shows a plateau in the same range of magnetic fields as the 
magnetization itself. These two observations strongly suggest that the 
system with PBC along the rungs has {\it nonmagnetic} excitations which
do not contribute to the magnetization or susceptibility, but do contribute
to the specific heat. Figure \ref{cvchi36} gives a more direct comparison
between OBC and PBC along the rungs. The LEH of Sec. III will clearly show the
origin of these excitations. Although these excitations were studied by
previous authors \cite{cabr1,cabr2,kawa}, we believe that our specific heat 
plots prove their existence most physically. To show these excitations even 
more explicitly, we present in Fig. \ref{nrg12} all the energy levels 
for a $12$-site chain in the sector $S^z =2$ (i.e., $m_s =1/2$) using exact 
diagonalization. It is clear that the ground state is well separated from 
the excited states for OBC, but it is at the bottom of a
band of excitations for PBC; these excitations are nonmagnetic since
they have the same value of $S^z$ as the ground state. 

We should point out that the rapid but small fluctuations seen in 
Figs. \ref{chio36} - 
\ref{cvchi36}, in the susceptibility and specific heat at the lowest 
temperature of $T=0.1$, are due to finite-size effects. Apart from a large 
plateau at $m_s =1/2$, a system with only $36$ sites also has small plateaus 
for several values of $m$ at zero temperature. These lead to small 
wiggles in the magnetization $\langle M \rangle$ at very low temperature. 
The wiggles get amplified in the susceptibility since it is equal to
the first derivative, i.e., $\chi = \partial \langle M \rangle /\partial h$.
The specific heat shows low-temperature fluctuations for the same reason.

We should mention here that a small plateau has been found at $m_s =0$ for PBC 
along the rungs \cite{cabr2,kawa}. The half-width is given by $h_{c+} /J = 
0.21$ in the limit $\jp /J \rightarrow \infty$. However, this plateau is not 
clearly visible in our low-temperature plots of magnetization and 
susceptibility.

\section{Low-energy effective Hamiltonians}

\subsection{General comments}

We will now discuss the LEH approach for studying 
the properties of spin ladders. There are two possible limits which may be 
considered. One could examine $\jp /J \rightarrow 0$ which corresponds to 
weakly interacting chains, and then directly use techniques from bosonization 
and conformal field theory; this has been done in detail by others
\cite{cabr1,cabr2,tots1,tots2}. We will therefore consider the strong-coupling
limit $J /\jp \rightarrow 0$ which corresponds to almost decoupled rungs. In
that limit, the LEH has been derived to first order in $J /\jp$ for a 
three-chain ladder with PBC along the rungs \cite{schu,kawa}, and for a 
two-chain ladder \cite{tots2,mila}. 

We will derive the LEH for the three-chain model with OBC along the rungs
and a two-chain model to {\it second order} in $J /\jp$, and for the 
three-chain model with PBC along the rungs to first order. For the 
three-chain system with OBC and for the 
two-chain system, we find that the first-order LEH takes the form of the 
$XXZ$ spin-$1/2$ model in a magnetic field. A lot of information is available 
for this model through conformal field theory \cite{cabr2,hald2}. In
particular, the exponent $\eta$ for the correlation power laws can be
read off from the first-order Hamiltonian. We will use the terms of
second order in $J / \jp$ only to determine the boundaries $h_{c \pm}$ of the 
various plateaus. The second-order terms should also 
give corrections to the exponent $\eta$ but we will not consider that problem
here. For the three-chain model with PBC along the rungs, even the first-order 
LEH is sufficiently complicated that its properties are not well 
understood; however we will present the form of the LEH for completeness.

We derive the LEH as follows. We first set the intrachain coupling $J=0$ and 
consider which of the states of a single rung are degenerate in energy in 
the presence of a magnetic field. In general, there will be several values 
of the field, denoted by $h_0$, for which two or more of the rung states will 
be degenerate ground states. We will consider each such value of $h_0$ in 
turn. The degenerate rung states will constitute our low-energy states. If 
the amount of degeneracy in each rung is $d$, the total
number of low-energy states in a system with $L$ rungs is given by $L^d$.
(The number $d$ depends both on the system and on the 
field $h_0$. It is two for three chains with OBC along the rungs and for two 
chains, while it is three or four for three chains with PBC along the rungs. 
The form of the LEH depends crucially on this degeneracy). Next, we decompose 
the Hamiltonian of the total system 
as $H= H_0 + V$, where $H_0$ contains only the rung interaction $\jp$ and
the field $h_0$, and $V$ contains the small interactions $J$ and the 
residual magnetic field $h - h_0$ which are both assumed to be much smaller 
than $\jp$. Let us now denote the degenerate and low-energy states of the 
system as $p_i$ and the high-energy states as $q_{\alpha}$. The low-energy
states all have energy $E_0$, while the high-energy states have energies 
$E_{\alpha}$ according to the exactly solvable Hamiltonian $H_0$. Then
the first-order LEH is given, up to an overall constant, by degenerate 
perturbation theory, 
\beq
H_{eff}^{(1)} ~=~ \sum_{ij} ~\vert p_i \rangle ~\langle p_i \vert V \vert p_j 
\rangle ~\langle p_j \vert ~.
\label{ham2}
\eeq
The second-order LEH is given by
\beq
H_{eff}^{(2)} ~=~ \sum_{ij} ~\sum_{\alpha} ~\vert p_i \rangle ~\frac{\langle 
p_i \vert V \vert q_{\alpha} \rangle ~\langle q_{\alpha} \vert V \vert 
p_j \rangle}{E_0 ~-~ E_{\alpha}} ~\langle p_j \vert ~.
\label{ham3}
\eeq
The calculation of the various matrix elements in Eqs. (\ref{ham2}) and 
(\ref{ham3}) can be simplified by using the symmetries of the 
perturbation $V$, e.g., translations and rotations about the $\hat z$ axis.

Finally, if there is a state $p_i$ such that $\langle p_j \vert V \vert p_i
\rangle =0$ for all low-energy states $j \ne i$, then the unnormalized state
$p_i$ is given, to first order, by
\beq
\vert p_i \rangle^{(1)} ~=~ \vert p_i \rangle ~+~ \sum_{\alpha} ~\vert
q_{\alpha} \rangle ~\frac{\langle q_{\alpha} \vert V \vert p_i \rangle}{E_0 ~
-~ E_{\alpha}} ~.
\label{pi}
\eeq
This result will be used to compute the first-order changes in some
quantities like the spin densities and correlations.

Before ending this section, we would like to make a few comments on
the $XXZ$ spin-$1/2$ model in a magnetic field since this will play an
important role below \cite{cabr2,hald2}. Consider a spin-$1/2$ chain governed 
by the Hamiltonian
\beq
H ~=~ \sum_n ~[~ S_n^x S_{n+1}^x ~+~ S_n^y S_{n+1}^y ~+~ \Delta ~S_n^z 
S_{n+1}^z ~]~ -~ h ~\sum_n ~S_n^z ~.
\label{hamxxz}
\eeq
where the anisotropy parameter $\Delta > -1$. It is known that this system 
is gapped for $h > 1+ \Delta$ with all sites having $S_z = 1/2$ in the 
ground state, and for $h < -1- \Delta$ with all sites having $S_z = -1/2$. 
For $\Delta \le 1$, these are the only two magnetization plateaus with 
$m = \pm 1/2$ per site. For $\Delta \le 1$ and $h=0$, the two-spin 
correlations decay asymptotically as
\bea
\langle S_0^+ S_n^- \rangle ~& \sim &~ \frac{(-1)^n}{\vert n \vert^\eta} ~,
\nonumber \\
\langle S_0^z S_n^z \rangle ~& \sim &~ \frac{(-1)^n}{\vert n \vert^{1/\eta}}~,
\nonumber \\
\eta ~&=&~ \frac{1}{2} ~+~ \frac{1}{\pi} ~\sin^{-1} ~(\Delta ) ~.
\label{corrxxz}
\eea
On the other hand, for $\Delta > 1$, there is an 
additional plateau at $m_s =0$; there are two
degenerate ground states which have a period of two sites consistent with
the condition (\ref{quant}). Thus the invariance of the Hamiltonian under a 
translation by one site is spontaneously broken in the ground states. 
This is particularly obvious for $\Delta \rightarrow \infty$ 
where the two ground states are $+-+- \cdots$ and $-+-+ \cdots$. The
two-spin correlations decay exponentially for $\Delta > 1$ and $h=0$.

\subsection{Three-chain ladder with open boundary condition along the 
rungs}

We will decompose the Hamiltonian in (\ref{ham1}) as $H= H_0 + V$, where
\bea
H_0 ~&=&~ \jp ~\sum_{a=1,2} ~\sum_n ~ {\bf S}_{a,n} \cdot {\bf S}_{a+1,n} ~
-~ h_0 ~\sum_{a=1}^3 ~\sum_n ~S_{a,n}^z ~, \nonumber \\
V ~&=&~  J ~\sum_{a=1}^3 ~\sum_n ~{\bf S}_{a,n} \cdot {\bf S}_{a,n+1} ~-~ 
(h ~-~ h_0 ) ~\sum_{a=1}^3 ~\sum_n ~S_{a,n}^z ~.
\label{ham4}
\eea
We determine the field $h_0$ by considering the rung Hamiltonian $H_0$ and 
identifying the values of the magnetic field $h_0$ where two or more of 
the rung states become degenerate. 

The eight states in each rung are described by specifying the $S^z$ 
components ($+$ and $-$ denoting $+1/2$ and $-1/2$ respectively) of the sites 
belonging to chains $1$, $2$ and $3$. For instance, the four states with 
total $S=3/2$ are denoted by $\vert 1 \rangle$, ..., $\vert 4 \rangle$, 
where $\vert 1 \rangle = \vert + + + \rangle$ and the other three states can be
obtained by acting on it successively with the operator $S^- = \sum_a S_a^-$.
These four states have the energy $\jp /2$ in the absence of a magnetic field. 
There is one doublet of states $\vert 5 \rangle$ and $\vert 6 \rangle$ 
with $S=1/2$, where $\vert 5 \rangle = [~2 ~\vert + - + \rangle - \vert - + + 
\rangle - \vert + + - \rangle ~]/ {\sqrt 6}$ and $\vert 6 \rangle \sim S^- 
\vert 5 \rangle$. These have energy $- \jp$. Finally, there is another 
doublet of states $\vert 7 \rangle = [~ \vert + + - \rangle - \vert - + + 
\rangle ~]/ {\sqrt 2}$ and $\vert 8 \rangle \sim S^- \vert 7 \rangle$ which 
have zero energy. It is now evident that the state $\vert 1 \rangle$
with $S^z = 3/2$ and the state $\vert 5 \rangle$ with $S^z =1/2$
become degenerate at a magnetic field $h_0 = 3\jp /2$,
while states $\vert 5 \rangle$ and $\vert 6 \rangle$ are trivially 
degenerate for the field $h_0 = 0$. We will now examine these two cases
separately.

For $h_0 = 3\jp /2$, the low-energy states in each rung are given by
$\vert 1 \rangle$ and $\vert 5 \rangle$, while the other six are 
high-energy states. We thus have 
an effective spin-$1/2$ object on each rung $n$. We may introduce three 
spin-$1/2$ operators $(S_n^x , S_n^y , S_n^z )$ for each rung such that
$S_n^{\pm} = S_n^x \pm i S_n^y$ and $S_n^z$ have the following actions:
\bea
S_n^+ ~\vert 1 \rangle_n ~&=&~ 0 ~, ~~~~ 
S_n^+ ~\vert 5 \rangle_n ~=~ \vert 1 \rangle_n ~, \nonumber \\
S_n^- ~\vert 1 \rangle_n ~&=&~ \vert 5 \rangle_n ~, ~~~~
S_n^- ~\vert 5 \rangle_n ~=~ 0 ~, \nonumber \\
S_n^z ~\vert 1 \rangle_n ~&=&~ \frac{1}{2} ~\vert 1 \rangle_n ~, ~~~~ 
S_n^z ~\vert 5 \rangle_n ~=~ -~\frac{1}{2} ~\vert 5 \rangle_n ~. 
\label{sn}
\eea
Note that the state which has a $\vert 1 \rangle$ on every rung, i.e., 
$\vert 111 \cdots \rangle$, is just the state with rung 
magnetization $m_s =3/2$ corresponding to the saturation plateau. The state 
with a $\vert 5 \rangle$ on every rung corresponds to
the $m_s =1/2$ magnetization plateau. The LEH we are trying to derive will
therefore describe the transition between these two plateaus.

We now turn on the perturbation $V$ in (\ref{ham4}) with the assumption that 
$J$ and $h - h_0$ are both much smaller than $\jp$. We can write $V = \sum_n
V_{n,n+1}$, where 
\beq
V_{n,n+1} ~=~ J ~\sum_{a=1}^3 ~{\bf S}_{a,n} \cdot {\bf S}_{a,n+1} ~-~ 
\frac{1}{2} (h ~-~ h_0 ) ~\sum_{a=1}^3 ~[~ S_{a,n}^z ~+~ S_{a,n+1}^z ~]~.
\eeq
The action of $V_{n,n+1}$ on the four low-energy states involving rungs 
$n$ and $n+1$ can be obtained after a long but straightforward calculation.
We then use Eq. (\ref{ham3}) and find that the LEH to second order 
in $J/ \jp$ is given, up to a constant, by
\bea
H_{eff} ~=~ & & J ~\sum_n ~[~ S_n^x S_{n+1}^x ~+~ S_n^y S_{n+1}^y ~+~ ( 
\frac{1}{2} ~-~ \frac{29J}{72\jp})~ S_n^z S_{n+1}^z ~] \nonumber \\
& & -~ \frac{5J^2}{18\jp} ~\sum_n ~(\frac{1}{2} - S_n^z)~(S_{n-1}^x S_{n+1}^x 
+ S_{n-1}^y S_{n+1}^y ) \nonumber \\
& & -~ \frac{2J^2}{27\jp} ~\sum_n ~(\frac{1}{2} - S_{n-1}^z )~
(\frac{1}{2} - S_n^z )~(\frac{1}{2} - S_{n+1}^z ) \nonumber \\
& & -~(~ h ~-~ \frac{3\jp}{2} ~-~ \frac{J}{2} ~-~ \frac{29J^2}{72\jp} ~)~
\sum_n ~S_n^z ~,
\label{ham5}
\eea
where we have substituted $h_0 = 3\jp /2$. Note that the terms of order $J$
only involve two neighboring sites. The LEH up to that order simply describes
an $XXZ$ model with anisotropy $\Delta =1/2$ in a magnetic field $h - 3\jp /2
- J/2$ [see (\ref{hamxxz})]. Some of the terms of order $J^2/ \jp$ involve 
three neighboring sites; this makes the model unsolvable by the Bethe ansatz 
at this order.

We will now use (\ref{ham5}) to compute the values of the fields $h_1$ and
$h_2$ where the states with all rungs equal to $\vert 1 \rangle$ and all 
rungs equal to $\vert 5 \rangle$ respectively become the ground states. 
We can then identify $h_1$ with the lower critical field $h_{c-}$ for the 
plateau at $m_s =3/2$, and $h_2$ with the upper critical field $h_{c+}$ for 
the plateau at $m_s =1/2$. [Recall the definition of upper and lower critical 
fields around Eqs. (\ref{hplus}) and (\ref{hminus})].

To compute the field $h_1$, we compare the energy $E_0$ of the state with 
all rungs equal to $\vert 1 \rangle $ with the minimum energy $E_{min} 
(k)$ of a spin-wave state in which one rung is equal to $\vert 5 \rangle$ 
and all the other rungs are equal to $\vert 1 \rangle$. A spin wave with 
momentum $k$ is given by
\beq
\vert k \rangle ~=~ \frac{1}{\sqrt L} ~\sum_n ~e^{ikn} ~\vert 5_n \rangle ~,
\eeq
where $\vert 5_n \rangle$ denotes a state where only the rung $n$ is 
equal to $\vert 5 \rangle$. 
The spin-wave dispersion, i.e., $\omega (k) = E (k) - E_0$, is found from 
(\ref{ham5}) to be
\beq
\omega (k) ~=~ J ~(~ \cos k ~-~ \frac{1}{2} ~+~ \frac{29J}{72 \jp} ~)~ 
+ ~(~ h ~-~ \frac{3\jp}{2} ~-~ \frac{J}{2} ~-~ \frac{29J^2}{\jp} ~) ~.
\eeq
This is minimum at $k =\pi$ and it turns negative there for $h < h_1$, where
\beq
h_1 ~=~ \frac{3\jp}{2} ~+~ 2 J ~.
\label{h1}
\eeq
This is therefore the transition point between the ferromagnetic state $\vert 
111 \cdots \rangle$ and a spin-wave band lying immediately below it in energy.

Similarly, we compute the field $h_2$ by comparing the energy $E_0$ of the
state with all rungs equal to $\vert 5 \rangle$ with the minimum energy 
$E_{min} (k)$ of a spin wave in which a $\vert 5 \rangle$ at one 
rung is replaced by a $\vert 1 \rangle$. For a spin wave with momentum $k$,
the dispersion $\omega (k) = E (k) - E_0$ is found to be
\beq
\omega (k) ~=~ J ~(~\cos k ~-~ \frac{1}{2} ~+~ \frac{29J}{72 \jp} ~)~
+ ~\frac{J^2}{\jp} ~(~ \frac{2}{9} ~-~ \frac{5}{18} ~\cos 2k ~) ~
- ~(~ h ~-~ \frac{3\jp}{2} ~-~ \frac{J}{2} ~-~ \frac{29J^2}{72 \jp} ~) ~.
\eeq
This is minimum at $k =\pi$ and it turns positive there for $h > h_2$, where
\beq
h_2 ~=~ \frac{3\jp}{2} ~-~ J ~+~ \frac{3J^2}{4\jp} ~.
\label{h2}
\eeq
This marks the transition between the state $\vert 555 \cdots \rangle$ and the 
spin-wave band. Equation (\ref{h2}) agrees to this order with the higher-order 
series given in the literature \cite{cabr2}. Note that the second-order 
result gives $h_2 /J = 3.75$ for $J /\jp =1/3$, compared to our DMRG 
value of $h_{c+} /J =3.79$ in (\ref{plat1}).

{}From the first-order terms in (\ref{ham5}), we can deduce the asymptotic
form of the two-spin correlations. From (\ref{corrxxz}), we see that the 
exponent $\eta = 2/3$ for $\Delta =1/2$. Although this is the exponent for 
the $+-$ correlation of the effective spin-$1/2$ defined on each rung,
we would expect the same exponent to appear in all the correlations
$\langle S_{a,l}^+ S_{b,n}^- \rangle$ studied by DMRG in the previous
section, regardless of how we choose the chain indices $a,b = 1,2,3$. We now
see that the analytically predicted exponent of $2/3$ agrees quite well with 
the numerically obtained exponents which lie in the range $0.61$ to $0.70$.
Incidentally, this agreement implies that $J /\jp = 1/3$ is sufficiently 
small so that the second-order terms do not significantly affect 
the correlation exponent.

Finally, we can use the first-order wave function given in (\ref{pi}) to 
compute the spin densities and short-distance two-spin correlations. As 
examples, we quote the results for spin densities and some of the 
nearest-neighbor spin correlations for the plateau at $m_s =1/2$. We will give 
the first-order expressions and their values for $J/\jp =1/3$, followed by 
the numerical values obtained by DMRG.
\bea
\langle S_{1,n}^z \rangle ~&=&~ \langle S_{3,n}^z \rangle ~=~ \frac{1}{3} ~
-~ \frac{4J}{27 \jp} ~=~ 0.28 ~~~ {\rm vs.} ~~~~~~ 0.27 ~~~~ {\rm from ~~
DMRG} ~, \nonumber \\
\langle S_{2,n}^z \rangle ~&=&~ - ~\frac{1}{6} ~+~ \frac{8J}{27 \jp} ~=~ 
-0.07 ~~~~~~~~~ {\rm vs.} ~~ -0.04 ~, \nonumber \\
\langle S_{1,n}^z S_{1,n+1}^z \rangle ~&=&~ \frac{1}{9} ~-~ \frac{J}{8\jp} ~
=~ 0.07 ~~~~~~~~~~~~~~~~~~ {\rm vs.} ~~~~~~ 0.07 ~, \nonumber \\
\langle S_{2,n}^z S_{2,n+1}^z \rangle ~&=&~ \frac{1}{36} ~-~ \frac{4J}{27\jp}~
=~ -0.02 ~~~~~~~~~~~~ {\rm vs.} ~~ -0.02 ~, \nonumber \\
\langle S_{1,n}^+ S_{1,n+1}^- \rangle ~&=&~ -~ \frac{J}{6\jp} ~
=~ -0.06 ~~~~~~~~~~~~~~~~~~ {\rm vs.} ~~ -0.08 ~, \nonumber \\
\langle S_{2,n}^+ S_{2,n+1}^- \rangle ~&=&~ -~ \frac{2J}{9\jp}~
=~ -0.07 ~~~~~~~~~~~~~~~~~~ {\rm vs.} ~~ -0.11 ~. 
\eea

We will now consider the LEH at the other magnetic field $h_0 =0$ where the
rung states $\vert 5 \rangle$ and $\vert 6 \rangle$ are degenerate. 
We take these as the low-energy states and introduce
new effective spin-$1/2$ operators for each rung with actions similar
to Eqs. (\ref{sn}), except that we replace $\vert 1 \rangle$ and $\vert 5 
\rangle$ in those equations by $\vert 5 \rangle$ and $\vert 6 \rangle$. 
We again compute the action of the perturbation $V$ on the low-energy states. 
We then deduce the second-order LEH to be
\beq
H_{eff} ~=~ J ~\sum_n ~[~ ( ~1 ~-~ \frac{J}{9\jp} ~)~ {\bf S}_n \cdot 
{\bf S}_{n+1} ~-~ \frac{8J}{27\jp} ~{\bf S}_n \cdot {\bf S}_{n+2} ~] ~-~ 
h ~\sum_n ~S_n^z ~.
\label{ham6}
\eeq
This Hamiltonian describes the transition between the magnetization plateaus 
at $m_s =1/2$ and $m_s =-1/2$; since these plateaus are reflections of each
other about zero magnetic field, it is sufficient to study one of them. By a 
calculation similar to the one used to derive (\ref{h1}), the field $h_1$ 
can be found from the dispersion of a spin wave in which one rung
is equal to $\vert 6 \rangle$ and all the other rungs are equal to $\vert 5
\rangle$. The dispersion is
\beq
\omega (k) ~=~ h ~+~ (~ J ~-~ \frac{J^2}{9\jp} ~)~(~ \cos k ~-~ 1 ~)~ +~ 
\frac{8J^2}{27\jp} ~(~ 1 ~-~ \cos 2k ~)~.
\eeq
This gives
\beq
h_1 ~=~ 2J ~-~ \frac{2J^2}{9\jp} ~.
\eeq
This is the lower critical field $h_{c-}$ of the $m_s =1/2$ plateau. For $J/
\jp = 1/3$, the second-order result gives $1.93$ versus the DMRG value of 
$1.91$ in (\ref{plat1}). The Hamiltonian (\ref{ham6}) describes an isotropic
spin-$1/2$ antiferromagnet with a weak ferromagnetic next-nearest-neighbor
interaction. From the comments at the end of the previous section, we see 
that this model only has the two saturation plateaus at $m_s =\pm 1/2$, and no 
other plateau in between. For $h=0$, the two-spin correlations decay as power 
laws with the exponent $\eta =1$ [see (\ref{corrxxz})].

\subsection{Three-chain ladder with periodic boundary condition along the
rungs}

In this section, we will present the first-order LEH for the Hamiltonian 
(\ref{ham1}) with PBC along the rungs. The LEH will turn out to be somewhat 
complicated.  We will not study their properties in any detail, but will limit 
ourselves to a few comments. As in the case with OBC along the rungs, there 
are two different LEH
to be considered here because there are two values of the magnetic field where 
there are degeneracies. We again begin with a description of the eight states 
on each rung. The four states with $S=3/2$ are the same as the states
$\vert 1 \rangle$, ...., $\vert 4 \rangle$ introduced in the previous section,
except that they now have energy $3\jp /4$ in the absence of a field. The 
doublet states have to be chosen differently now in order that they be 
eigenstates of the periodic rung Hamiltonian. We choose two of the doublet 
states to be $\vert 5^\prime \rangle = [ ~\vert + + - \rangle + \omega^2 ~\vert 
+ - + \rangle + \omega ~\vert - + + \rangle ~]/ {\sqrt 3}$ and $\vert 6^\prime 
\rangle \sim S^- \vert 5^\prime \rangle$,
where $\omega = \exp (i2\pi /3)$. These two states have momenta $2\pi /3$ along
the rung (right moving). The other two doublet states are $\vert 7^\prime 
\rangle = [ ~\vert + + - \rangle + \omega ~\vert + - + \rangle + \omega^2 ~
\vert - + + \rangle ~]/ {\sqrt 3}$ and $\vert 8^\prime \rangle \sim S^- \vert 
7^\prime \rangle$ with momenta $-2\pi /3$ (left moving). All these four states 
have energy $-3 \jp /4$. This extra degeneracy (which is twice the degeneracy 
of the doublets for OBC along the rungs) leads to a more complicated LEH as 
we will see.

We now note that for a field $h_0 =3 \jp /2$, the three states $\vert 1 
\rangle$, $\vert 5^\prime \rangle$ and $\vert 7^\prime \rangle$ become 
degenerate. We now introduce seven operators $R^{\pm}, L^{\pm}, \tau^{\pm}$
and $\sigma^z$ for each rung with the 
following nonzero actions on the three low-energy states,
\bea
R^+_n ~\vert 5^\prime \rangle_n ~&=&~ \vert 1 \rangle_n ~, ~~~~
R^-_n ~\vert 1 \rangle_n ~=~ \vert 5^\prime \rangle_n ~, \nonumber \\
L^+_n ~\vert 7^\prime \rangle_n ~&=&~ \vert 1 \rangle_n ~, ~~~~
L^-_n ~\vert 1 \rangle_n ~=~ \vert 7^\prime \rangle_n ~, \nonumber \\
\tau^+_n ~\vert 7^\prime \rangle_n ~&=&~ \vert 5^\prime \rangle_n ~, ~~~~
\tau^-_n ~\vert 5^\prime \rangle_n ~=~ \vert 7^\prime \rangle_n ~, 
\nonumber \\
\sigma^z_n ~\vert 1 \rangle_n ~&=&~ \vert 1 \rangle_n ~, ~~~~
\sigma^z_n ~\vert 5^\prime \rangle_n ~=~ -~ \vert 5^\prime \rangle_n ~, ~~~~
\sigma^z_n ~\vert 7^\prime \rangle_n ~=~ -~ \vert 7^\prime \rangle_n ~. 
\label{snp}
\eea
All the actions (of operators on states) not mentioned in 
Eqs. (\ref{snp}) are assumed to give zero.
We thus observe that there are five magnetic operators $L^{\pm}, R^{\pm}$ and
$\sigma^z$ which change or measure the $S^z$ of a state, and two nonmagnetic
operators $\tau^{\pm}$ which do not change $S^z$ but simply interchange
the right and left moving states.

We then find that the first-order LEH is given, up to a constant, by
\bea
H_{eff} ~=~ & & \frac{J}{2}~ \sum_n ~[~ L^+_n L^-_{n+1} ~+~ L^-_n L^+_{n+1} ~
+~ R^+_n R^-_{n+1} ~+~ R^-_n R^+_{n+1} ~]~ \nonumber \\
& & +~ \frac{J}{3} ~\sum_n ~[~ \tau^+_n \tau^-_{n+1} ~+~ \tau^-_n 
\tau^+_{n+1} ~ ]~ +~ \frac{J}{12} ~\sum_n ~\sigma^z_n \sigma^z_{n+1} 
\nonumber \\
& & -~\frac{1}{2}~ (~ h ~-~ \frac{3\jp}{2} ~-~ \frac{2J}{3} ~) ~\sum_n ~
\sigma^z_n ~.
\label{ham7}
\eea
We can now find the magnetic field $h_1$ at which the the ferromagnetic 
state $\vert 111 \cdots \rangle$ crosses over to the minimum energy spin-wave 
state (in which a $\vert 1 \rangle$ is replaced by either a $\vert 
5^\prime \rangle$ or a $\vert 7^\prime \rangle$ on exactly one rung). The 
spin-wave dispersion is
\beq
\omega (k) ~=~ h ~-~ \frac{3\jp}{2} ~-~ J ~+~ J ~\cos k ~.
\eeq
We thus see that $h_1 = 3 \jp /2 + 2J$ just as for OBC along the rungs. 
Thus the lower critical field $h_{c-}$ of the saturation plateau 
$m_s =3/2$ has the same value for OBC and PBC along the rungs. 

Below some field $h_2$ (which seems rather hard to find analytically), the 
low-energy eigenstates of (\ref{ham7}) will not
have the state $\vert 1 \rangle$ on any rung; only the states $\vert 
5^\prime \rangle$ and $\vert 7^\prime \rangle$ will appear. This gives us the
magnetization plateau $m_s =1/2$. However, this plateau has a large 
number of {\it nonmagnetic} excitations described by the Hamiltonian
\beq
H_{eff} ~=~ \frac{J}{3} ~\sum_n ~[~ \tau^+_n \tau^-_{n+1} ~+~ \tau^-_n 
\tau^+_{n+1} ~] 
\label{ham8}
\eeq
which may be obtained from (\ref{ham7}) by omitting 
the state $\vert 1 \rangle$ on all the rungs. 
Equation (\ref{ham8}) has the form of (\ref{hamxxz}) with $\Delta =0$, and is
therefore exactly solvable; at low temperature, it has a specific heat which 
grows linearly with $T$. The situation is therefore quite different from 
the case of OBC along the rungs where the $m_s =1/2$ plateau consists of a 
single state in which every rung is in the state $\vert 5 \rangle$; all other 
states are separated by a gap, hence the specific heat goes to zero 
exponentially at low temperature.

Finally, we examine the LEH at the field $h_0 =0$ where the four doublet 
states $\vert 5^\prime \rangle , \cdots, \vert 8^\prime \rangle$ become
degenerate. This has been discussed in detail earlier \cite{cabr2,schu,kawa}.
On each rung, we introduce effective spin-$1/2$ operators which change
or measure $S^z$, and the two nonmagnetic operators $\tau^{\pm}$ which 
interchange the left and right moving states. Then the LEH is
\beq
H_{eff} ~=~ \frac{J}{3} ~\sum_n ~[~ 1 ~+~ 4 ~(~\tau^+_n \tau^-_{n+1} ~+~ 
\tau^-_n \tau^+_{n+1} ~) ~]~ {\bf S}_n \cdot {\bf S}_{n+1} ~-~ h~ 
\sum_n ~S_n^z ~.
\label{ham9}
\eeq
This also appears to be nonexactly solvable but it has been studied 
numerically \cite{cabr2,kawa}. It has a small plateau at $m_s =0$ where there 
are two degenerate ground states, each with a period of two rungs. Above some 
magnetic field $h_1$ (which is again hard to calculate analytically from 
(\ref{ham9}), this model crosses over to the $m_s =1/2$ plateau where the rungs 
can only be in the two $S^z =1/2$ states $\vert 5^\prime \rangle$ and $\vert 
7^\prime \rangle$. We see from (\ref{ham9}) that these two states are again 
governed by the Hamiltonian in (\ref{ham8}).

The phase diagrams of the three-chain ladder for OBC and PBC along the 
rungs are shown as functions of $\jp /J$ and $h/J$ in Figs. 6 (b) and (c) 
in Ref. [11]. We observe that the plateaus with $m_s = 1/2$ and $m_s = 3/2$ 
(called $M=1/3$ and $M=1$, respectively, in Ref. [11]) have large regions of 
stability for both OBC and PBC. The plateau with $m_s =0$ exists only 
in the case of PBC along the rungs, and it has a small region of stability 
close to $h/J =0$.

\subsection{A two-chain ladder}

In this section, we will use the LEH approach to study a two-chain spin-$1/2$
ladder with the following Hamiltonian,
\bea
H ~=~ & & \jp ~\sum_n ~ {\bf S}_{1,n} \cdot {\bf S}_{2,n} ~+~ J_2 ~
\sum_{a=1}^2 ~\sum_n ~{\bf S}_{a,n} \cdot {\bf S}_{a,n+1} \nonumber \\ 
& & +~ 2 J_1 ~\sum_n ~{\bf S}_{1,n} \cdot {\bf S}_{2,n+1} ~-~ h ~
\sum_{a=1}^2 ~\sum_n ~S_{a,n}^z ~,
\label{ham10}
\eea
as shown in Fig. \ref{ladd2}. The model may be viewed as a single chain with 
an alternation in nearest-neighbor couplings $\jp$ and $2J_1$ (dimerization), 
and a next-nearest-neighbor coupling $J_2$ (frustration). Equation 
(\ref{ham10}) has been studied extensively \cite{chit2,hamm}. More recently, 
it has been studied from the point of view of magnetization plateaus 
using a first-order LEH, bosonization and exact
diagonalization \cite{tots2,tone,mila} . We will therefore limit 
ourselves to deriving the second-order LEH and making a few other comments.

We begin by setting $J_1 = J_2 =0$, and studying the four states on each rung.
These are specified by giving the configurations $\pm$ of the spins on
chains $1$ and $2$ as follows. The three triplet states with $S=1$ are
denoted as $\vert 1 \rangle$, $\vert 2 \rangle$ and $\vert 3 \rangle$, where
$\vert 1 \rangle = \vert + + \rangle$ and the other two states are obtained
by acting on it successively with $S^-$. These three states
have energy $\jp /4$ in the absence of a magnetic field. The singlet state
$\vert 4 \rangle = [ \vert + - \rangle - \vert - + \rangle ]/ {\sqrt 2}$
has energy $-3\jp /4$. The states $\vert 1 \rangle$ and $\vert 4 \rangle$ 
become degenerate at a field $h_0 = \jp$. We now develop perturbation
theory by assuming that $J_1 , J_2$ and $h - h_0$ are all much less than
$\jp$. The perturbation is $V = \sum_n V_{n,n+1}$ where
\bea
V_{n,n+1} ~=~ & & J_2 ~\sum_{a=1}^2 ~ {\bf S}_{a,n} \cdot {\bf S}_{a,n+1} ~+~
2 J_1 ~{\bf S}_{1,n} \cdot {\bf S}_{2,n+1} \nonumber \\
& & -~\frac{1}{2} (h ~-~ h_0 ) ~\sum_{a=1}^2 ~[~S_{a,n}^z ~+~ S_{a,n+1}^z ~]~.
\eea
The actions of this operator on the four low-energy states of a pair of 
neighboring rungs can be easily obtained. We now introduce effective 
spin-$1/2$ operators ${\bf S}_n$ on each rung which act on the two low-energy 
states. The second-order LEH is then found to be
\bea
H_{eff} ~= & & (~ J_2 - J_1 - \frac{J_1^2}{\jp} ~) ~\sum_n ( S_n^x S_{n+1}^x +
S_n^y S_{n+1}^y ) \nonumber \\
& & + \frac{1}{2} (~ J_2 + J_1 + \frac{2J_1^2}{\jp} - \frac{3(J_1 - J_2)^2}{4
\jp} ~) ~\sum_n S_n^z S_{n+1}^z \nonumber \\
& & + \frac{J_1^2}{2\jp} \sum_n ~[ (\frac{1}{2} + S_n^z ) ( S_{n-1}^x 
S_{n+1}^x + S_{n-1}^y S_{n+1}^y ) + (\frac{1}{2} + S_{n-1}^z ) (S_n^x 
S_{n+1}^x + S_n^y S_{n+1}^y ) \nonumber \\
& & ~~~~~~~~~~~~~ + (\frac{1}{2} + S_{n+1}^z ) ( S_{n-1}^x S_n^x + 
S_{n-1}^y S_n^y ) + (\frac{1}{2} + S_{n-1}^z ) (\frac{1}{2} - 
S_n^z ) (\frac{1}{2} + S_{n+1}^z ) ] \nonumber \\
& & - \frac{(J_1 - J_2 )^2}{4 \jp} \sum_n ~(\frac{1}{2} - S_n^z ) 
( S_{n-1}^x S_{n+1}^x + S_{n-1}^y S_{n+1}^y ) \nonumber \\
& & -~ (~ h ~-~ \jp ~-~ \frac{J_1}{2} ~-~ \frac{J_2}{2} ~-~ \frac{3(J_1 - 
J_2 )^2}{8\jp} ~) ~\sum_n ~S_n^z ~.
\label{ham11}
\eea

We now compute the field $h_1$ above which the state $\vert 111 \cdots
\rangle$ becomes the ground state. The dispersion of a spin wave, in which 
one rung is equal to $\vert 4 \rangle$ and all the others are equal to $\vert
1 \rangle$, is given by 
\beq
\omega (k) ~=~ h ~-~ \jp ~-~ J_1 ~-~ J_2 ~-~ \frac{J_1^2}{2\jp} ~+~ (J_2 -
J_1 ) ~\cos k ~+~ \frac{J_1^2}{2\jp} ~\cos 2k ~.
\eeq
By minimizing this as a function of $k$ in various regions in 
the parameter space $(J_1 , J_2 )$, and then setting that minimum value equal
to zero, we find that $h_1$ is given by 
\bea
h_1 ~&=&~ \jp ~+~ 2 J_1 ~~{\rm if}~~ J_2 \le J_1 ~-~ \frac{2J_1^2}{\jp} ~, 
\nonumber \\
&=&~ \jp ~+~ J_1 ~+~ J_2 ~+~ \frac{J_1^2}{\jp} ~+~ \frac{(J_1 - J_2 )^2 
\jp}{4 J_1^2} ~~{\rm if}~~ J_1 - \frac{2J_1^2}{\jp} \le J_2 \le  ~J_1 ~+~ 
\frac{2J_1^2}{\jp} ~, \nonumber \\
&=&~ \jp ~+~ 2 J_2 ~~{\rm if}~~ J_2 \ge J_1 ~+~ \frac{2J_1^2}{\jp} ~.
\eea
This is the lower critical field $h_{c-}$ of the saturation plateau with
magnetization $m_s =1$ per rung. Similarly, we can find the field $h_2$ from 
the dispersion of a spin wave in which one rung is equal to $\vert 1 \rangle$ 
and the rest are equal to $\vert 4 \rangle$. The dispersion is given by 
\beq
\omega (k) ~=~ -~ h ~+~ \jp ~+~ \frac{3(J_1 - J_2 )^2}{4\jp} ~-~ 
\frac{J_1^2}{\jp} ~+~ (J_2 - J_1 - \frac{J_1^2}{\jp} ) ~\cos k ~-~ 
\frac{(J_1 - J_2 )^2}{4\jp} ~\cos 2k ~.
\eeq
By setting the minimum of this equal to zero, we find that $h_2$ is given by 
\bea
h_2 ~&=&~ \jp ~+~ \frac{(J_1 ~-~ J_2 )^2}{2\jp} ~+~ J_2 ~-~ J_1 ~-~ 
\frac{2 J_1^2}{\jp} ~~ {\rm if} ~~ J_2 \le J_1 ~+~ \frac{J_1^2}{\jp} ~, 
\nonumber \\
&=&~ \jp ~+~ \frac{(J_1 ~-~ J_2 )^2}{2\jp} ~-~ J_2 ~+~ J_1 ~~ {\rm if} ~~ 
J_2 \ge J_1 ~+~ \frac{J_1^2}{\jp} ~.
\eea
This is the upper critical field $h_{c+}$ of the saturation plateau with
magnetization $m_s =0$ per rung.

Finally, we can see that the first-order terms in (\ref{ham11}) are of the
same form as the $XXZ$ model in (\ref{hamxxz}). We can always make the
coefficient of the first term in (\ref{ham11}) positive, if necessary 
by performing a rotation $S_n^x \rightarrow (-1)^n S_n^x$, $S_n^y 
\rightarrow (-1)^n S_n^y$ and $S_n^z \rightarrow S_n^z$. We then get a 
first-order Hamiltonian of the form
\bea
H_{eff} ~=~ & & \vert J_2 ~-~ J_1 \vert ~\sum_n ~[~ S_n^x S_{n+1}^x ~+~ 
S_n^y S_{n+1}^y ~]~ +~ \frac{1}{2} ~(~ J_2 ~+~ J_1 ~)~ \sum_n ~S_n^z 
S_{n+1}^z \nonumber \\
& & -~ (~ h ~-~ \jp ~-~ \frac{J_1}{2} ~-~ \frac{J_2}{2} ~)~ \sum_n ~S_n^z ~.
\label{ham12}
\eea
This is an $XXZ$ model with 
\beq
\Delta ~=~ \frac{J_2 ~+~ J_1}{2 ~\vert J_2 ~-~ J_1 \vert} ~.
\eeq
{}From the comments in Sec. III A, we therefore see that the two-chain ladder 
will have an additional plateau at $m_s =1/2$ for $\Delta > 1$, i.e., if $J_2 
+ J_1 > 2 \vert J_2 - J_1 \vert$. In particular, $\Delta = \infty$ for $J_2 
= J_1$ (this is called the Shastry-Sutherland line \cite{shas}); the $m_s 
=1/2$ plateau will then stretch all the way from the upper 
critical field of the $m_s =0$ plateau to the lower critical field of the 
$m_s =1$ plateau. This can be seen in Fig. \ref{tone} which is taken from Ref. 
[16]; the dimerization parameter $\alpha$ in that figure is related to our 
couplings by $\jp = 1 + \alpha$ and $2J_1 = 1 - \alpha$. Note that the 
$m_s =1/2$ plateau is particularly broad at $\alpha = 0.6$, i.e., $J_2 = J_1 
= 0.2$, and that it actually touches the $m_s =1$ plateau on the right. The 
fact that it does not extend all the way up to the $m_s =0$ plateau on the left 
is probably because we have ignored the second-order terms in (\ref{ham11}) 
which lead to deviations from the $XXZ$ model.

\section{Summary and Outlook}

We studied a three-chain spin-$1/2$ ladder with a large ratio of interchain
coupling to intrachain coupling using the DMRG method and a LEH approach.
For both OBC and PBC along the rungs, we found a wide plateau
with rung magnetization given by $m_s =1/2$. For the case of OBC, the two-spin
correlations are extremely short-ranged, and the magnetic susceptibility 
and specific heat are very small at low temperature in the plateau. All 
these are consistent with the large magnetic gap. At other values of $m$, the 
two-spin correlations fall off as power laws; the exponents can be found by 
using the first-order LEH which takes the form of an $XXZ$ model in a 
longitudinal magnetic field. For the case of PBC, the magnetic susceptibility 
is again very small at low temperature in the plateau. However the specific 
heat goes to zero much more slowly which dramatically shows the presence of 
nonmagnetic excitations. This can be understood from the LEH in (\ref{ham8}) 
which is an $XY$ model. Finally, we used the LEH approach to study a two-chain 
ladder with an additional diagonal interaction. In addition to a plateau at 
$m_s =0$, this system also has a plateau at $m_s =1/2$ for certain regions in 
parameter space. The $m_s =1/2$ plateau is interesting because it corresponds
to degenerate ground states which spontaneously break the translation
invariance of the Hamiltonian. This can be understood from the LEH which, at
first-order, is an $XXZ$ model with $\Delta > 1$.

An interesting problem for the future may be to take the second-order 
terms in the LEH presented in Secs. III B and III D, and to compute the 
corrections produced by them in the exponents of the correlation power laws.
This would require us to study the effects of a perturbation to the $XXZ$
spin-$1/2$ chain. This may not be difficult to do analytically since the
$XXZ$ model is integrable and exactly solvable by the Bethe ansatz. 

The quantization condition for magnetization given in (\ref{quant})
is reminiscent of the quantum Hall effect where the Hall conductivity shows 
plateaus as a function of the magnetic field \cite{pran}. However, it is
not clear if the magnetization quantization is as insensitive to disorder as
the conductivity quantization is known to be. Although a magnetization 
plateau may be expected to
survive small amounts of disorder (e.g., if the disorder strength
is much smaller than the energy gap), there seems to be no fundamental
physical principle, analogous to gauge invariance in the quantum Hall system,  
why the {\it value} of the magnetization should remain fixed at a simple 
rational value. In fact, the derivation of (\ref{quant}) assumes translation
invariance of the Hamiltonian which is certainly broken by disorder. It would 
therefore be interesting to study this issue, for instance,
by allowing a small amount of disorder in the couplings of the spin
ladder models discussed in this paper.

\vskip .7 true cm
\noindent {\bf Acknowledgments}

We thank Andreas Honecker and Sriram Shastry for useful discussions. 
We are grateful to T. Tonegawa 
for giving us permission to reproduce Fig. \ref{tone} from Ref. [16]. The 
present work has been partly supported by the Indo-French Centre for the 
Promotion of Advanced Research through project No. 1308-4, "Chemistry and 
Physics of Molecule based Materials".

\newpage

\noindent {\bf Figure Captions}
\vskip .5 true cm

\noindent
{1.} The energy/site in units of $J$ vs $1/N$ at the $m_s =1/2$ plateau, for 
$J/ \jp =1/3$. The curves indicate quadratic fits for (a) $E_0 (M+1,N)$, 
(b) $E_0 (M,N)$, and (c) $E_0 (M-1,N)$.

\noindent
{2.} Plateau widths vs $1/N$ for (a) $m_s =1/2$, (b) $m_s =0$, and (c) $m_s 
=1$. 

\noindent
{3.} Spin densities at the $m_s =1/2$ plateau for $J /\jp =1/3$. The upper
points (circles) denote the top chain $a=1$, while the lower points (triangles)
denote the middle chain $a=2$. $n=1$ and $20$ denote the end and middle rungs 
respectively.

\noindent
{4.} Correlation function $<S^+_{2,l}S^-_{2,n}>$ at the $m_s =1/2$ plateau for 
$J/\jp =1/3$.

\noindent
{5.} Correlation function $<S^z_{1,l}S^z_{1,n}>$ at the $m_s =1/2$ plateau.

\noindent
{6.} Correlation functions $<S^+_{1,l}S^-_{2,n}>$ in the $m_s =1$ state for 
$J/\jp =1/3$.

\noindent
{7.} Correlation function $<S^z_{2,l}S^z_{2,n}>$ in the $m_s =1$ state.

\noindent
{8.} Spin densities in the $m_s =1$ state for $J/ \jp =1/3$. The upper
points (circles) denote the top chain $a=1$, while the lower points (triangles)
denote the middle chain $a=2$. $n=1$ and $20$ denote the end and middle rungs 
respectively.

\noindent
{9.} Magnetization vs magnetic field for 36 sites, with OBC along rungs for 
$J/ \jp =1/3$.

\noindent
{10.} Magnetization vs magnetic field for 36 sites, with PBC along rungs for 
$J/ \jp =1/3$.

\noindent
{11.} Susceptibility vs magnetic field for 36 sites, with OBC along rungs.

\noindent
{12.} Specific heat in units of $k_B$ vs magnetic field for 36 sites, with OBC 
along rungs for $J /\jp = 1/3$.

\noindent
{13.} Specific heat in units of $k_B$ for 36 sites, with PBC along rungs.

\noindent
{14.} Comparisons of specific heat and susceptibility of the 36-site systems 
with OBC and PBC along the rungs.

\noindent
{15.} Comparison of the energy spectra in units of $J$ of the $12$-site 
system with OBC and PBC along the rungs. The energies in the $S_z = 2$ 
sector are shown for $J/ \jp =1/3$.

\noindent
{16.} Schematic diagram of the two-chain ladder with an additional diagonal 
interaction. The labels $1$ and $2$ denote sites in the upper and lower chains 
respectively.

\noindent
{17.} Phase diagram of the two-chain ladder as a function of $h$ and $\alpha$ 
for $J_2 =0.2$. In our notation, $\jp = 1 + \alpha$ and $2 J_1 = 1 - \alpha$. 
The numbers $0$, $1/2$ and $1$ in the figure correspond to the values of $m_s$
at the plateaus. Reproduced with permission from Ref. [16].

\newpage

\begin{figure}[ht]
\begin{center}
\epsfig{figure=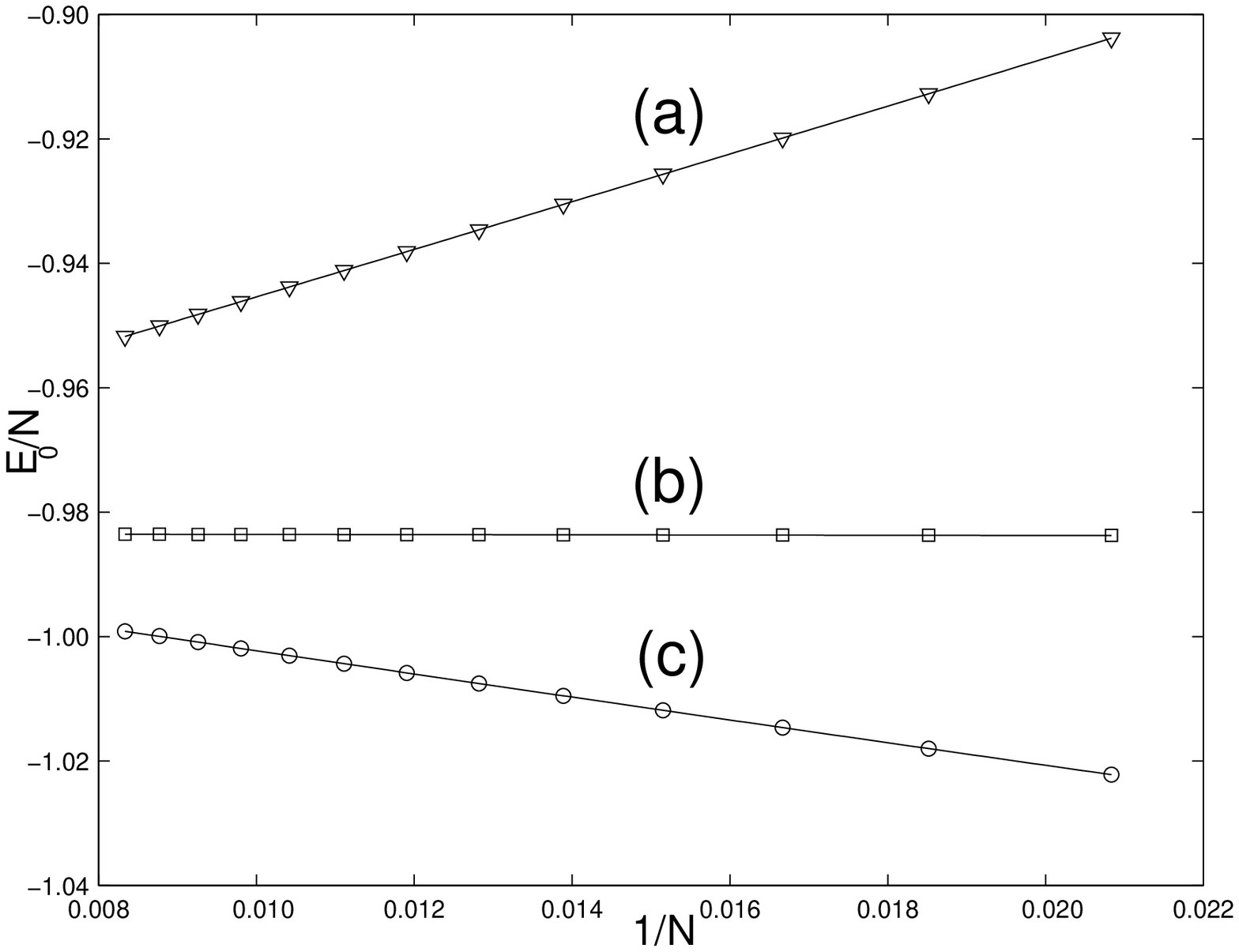,width=10cm}
\end{center}
\caption{}
\label{quadfit}
\end{figure}

\vspace*{1cm}
\begin{figure}[hp]
\begin{center}
\epsfig{figure=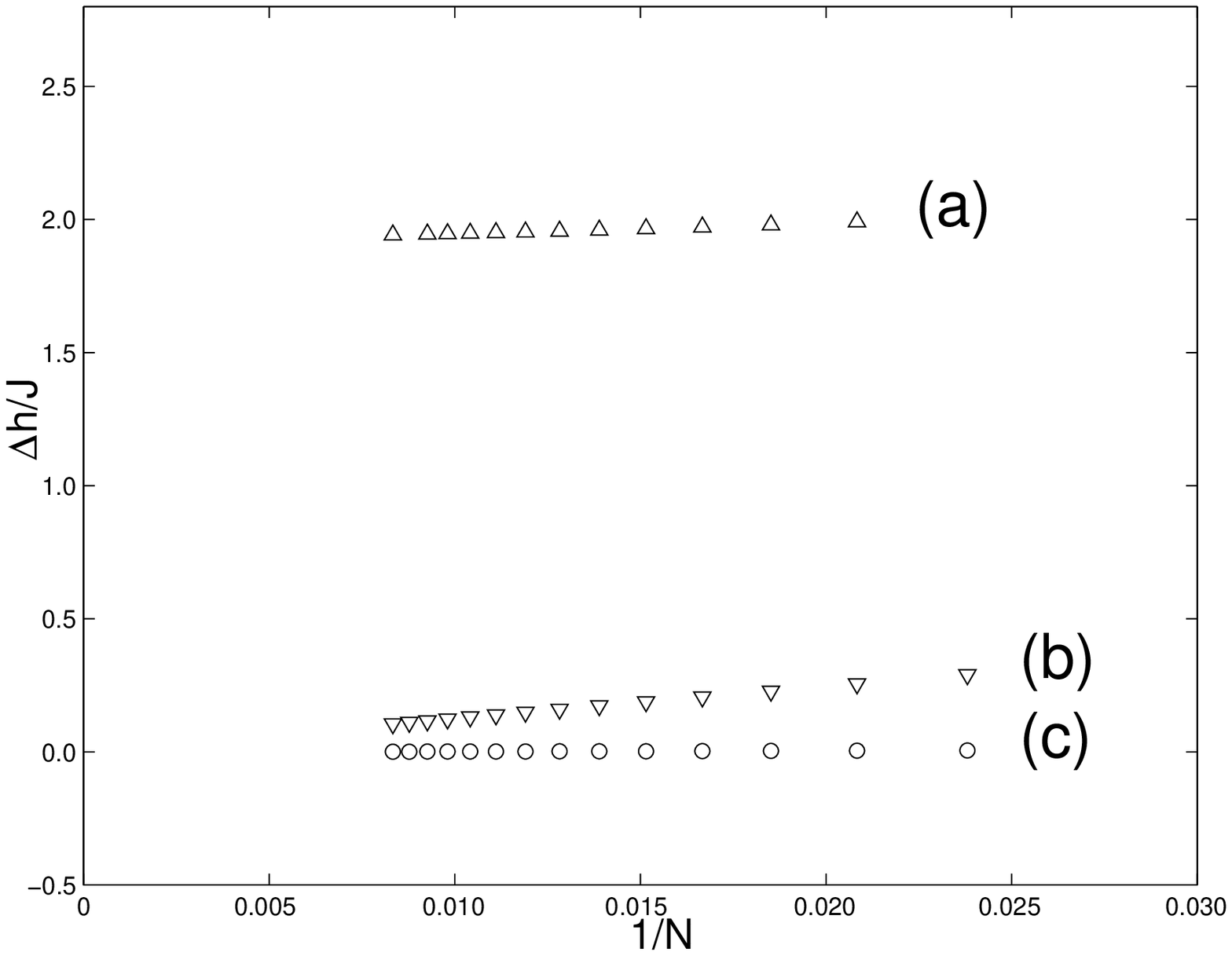,width=9cm}
\end{center}
\caption{}
\label{platgap}
\end{figure}

\begin{figure}[hp]
\begin{center}
\epsfig{figure=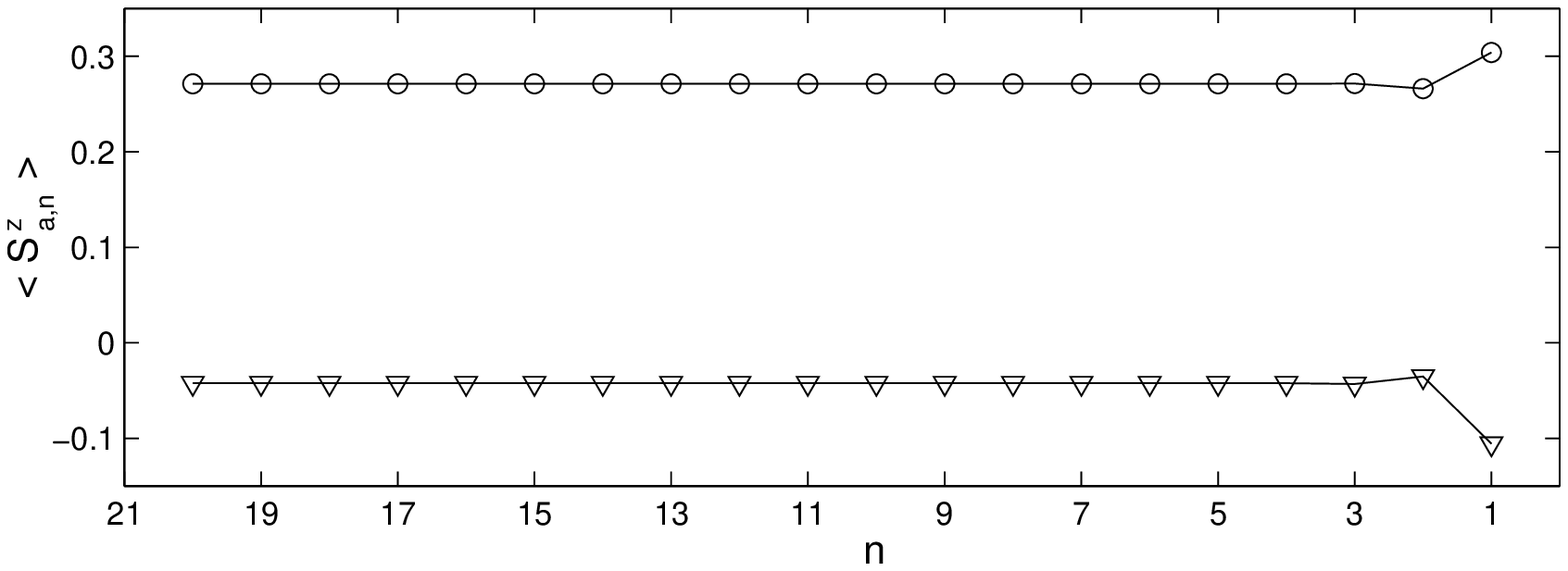,width=12cm}
\end{center}
\caption{}
\label{spnden1}
\end{figure}

\vspace*{1cm}
\begin{figure}[hp]
\begin{center}
\epsfig{figure=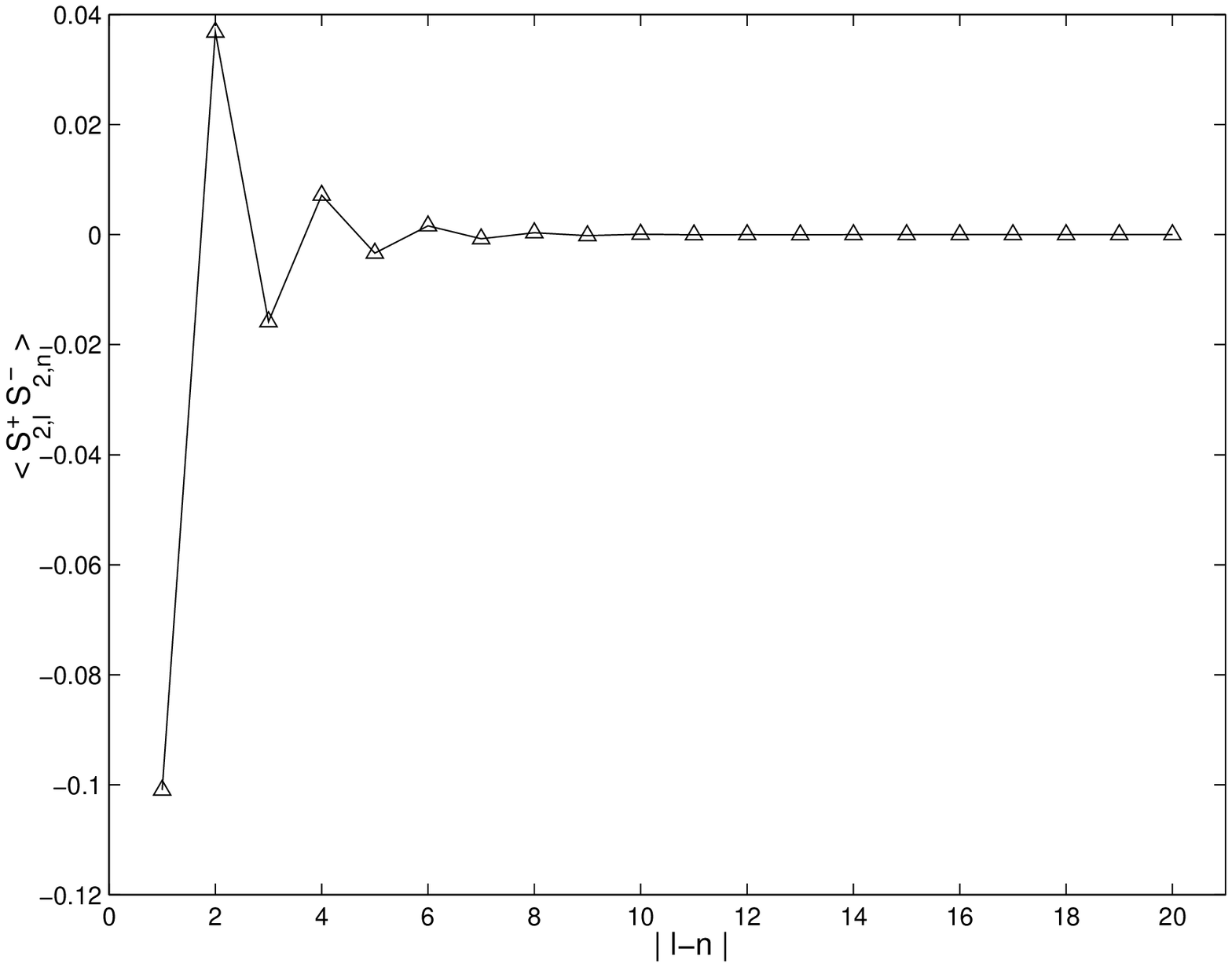,width=10cm}
\end{center}
\caption{}
\label{corr22pm}
\end{figure}

\begin{figure}[hp]
\begin{center}
\epsfig{figure=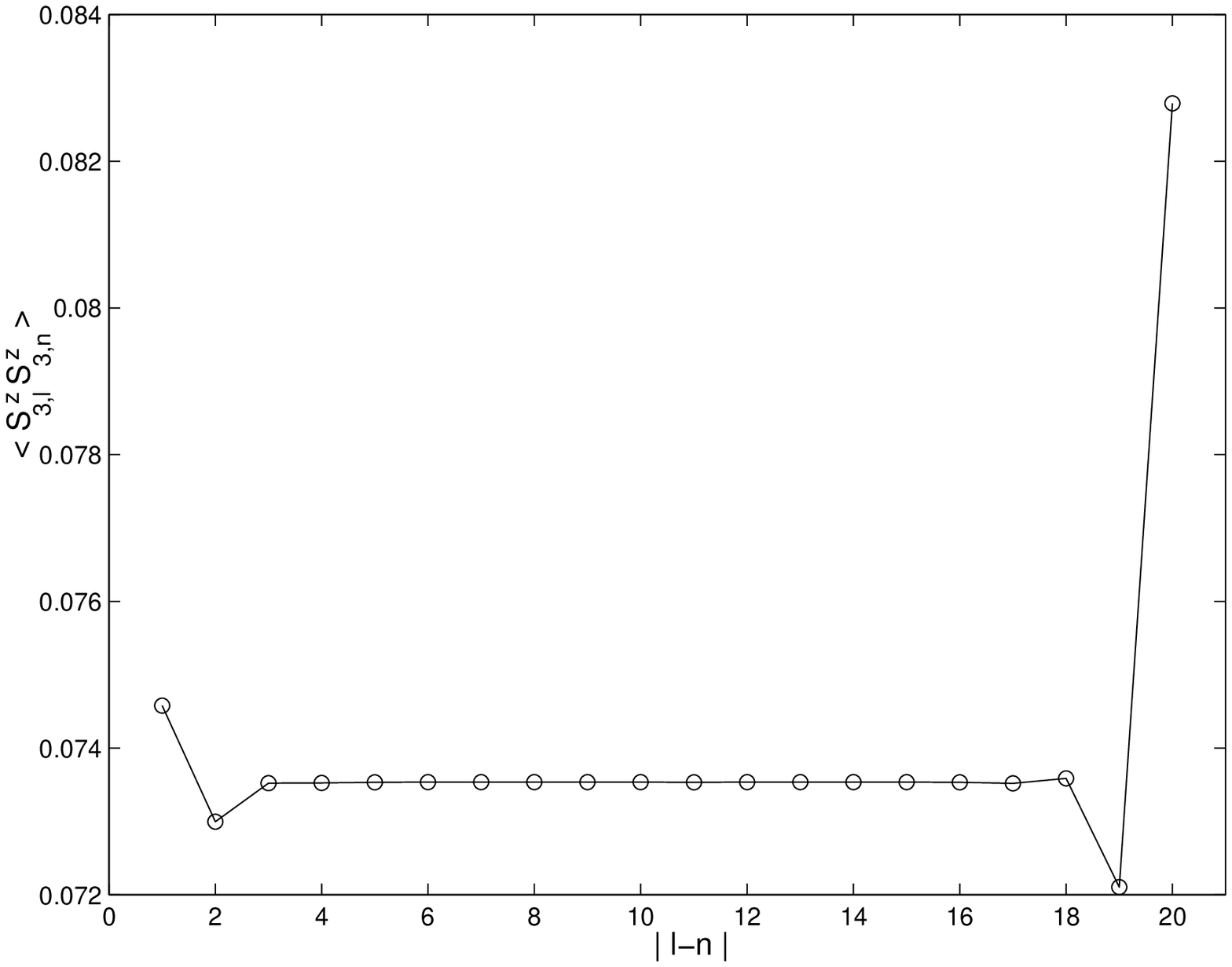,width=10cm}
\end{center}
\caption{}
\label{corr11zz}
\end{figure}

\vspace*{1cm}
\begin{figure}[hp]
\begin{center}
\epsfig{figure=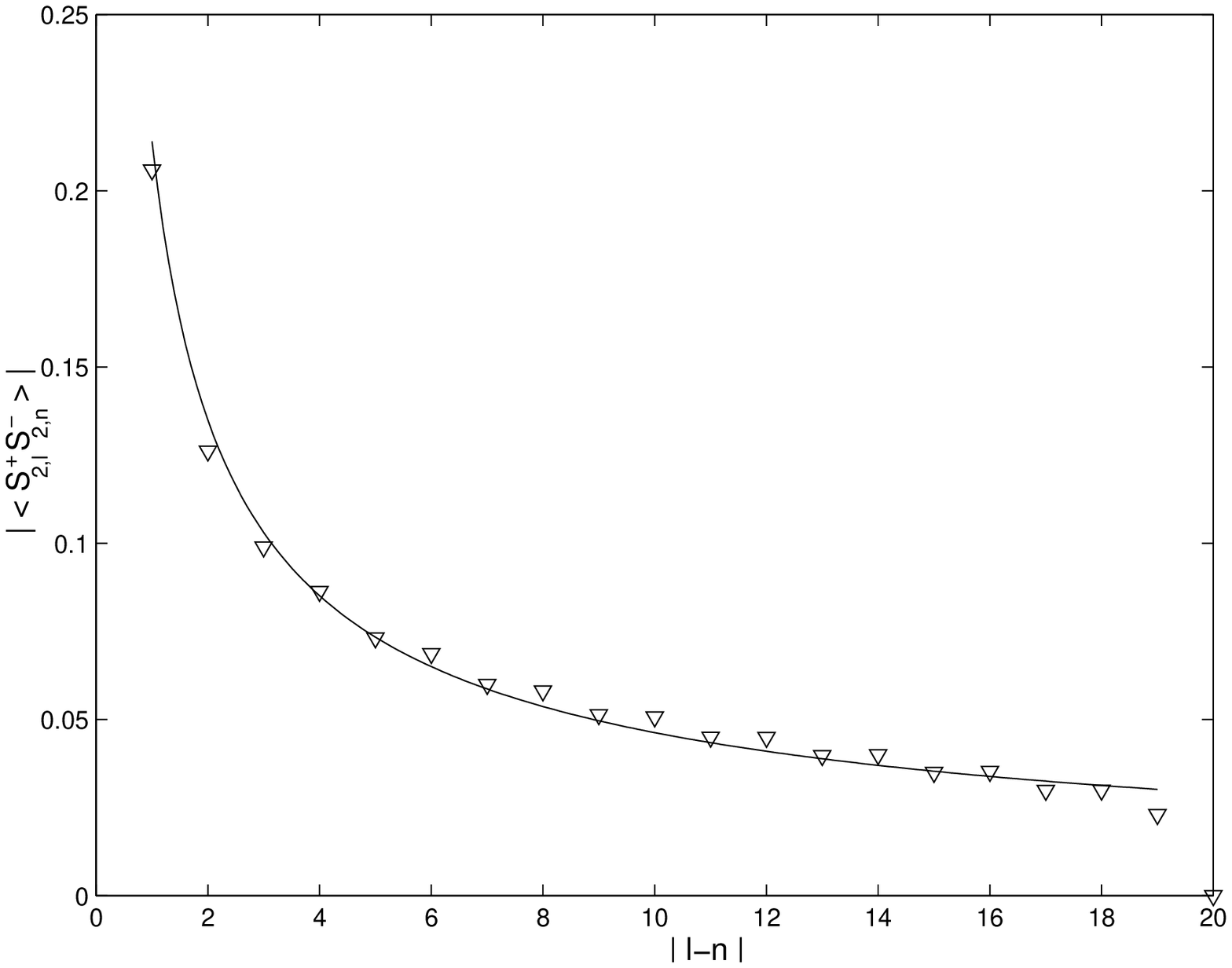,width=10cm}
\end{center}
\caption{}
\label{corr12pm}
\end{figure}

\begin{figure}[hp]
\begin{center}
\epsfig{figure=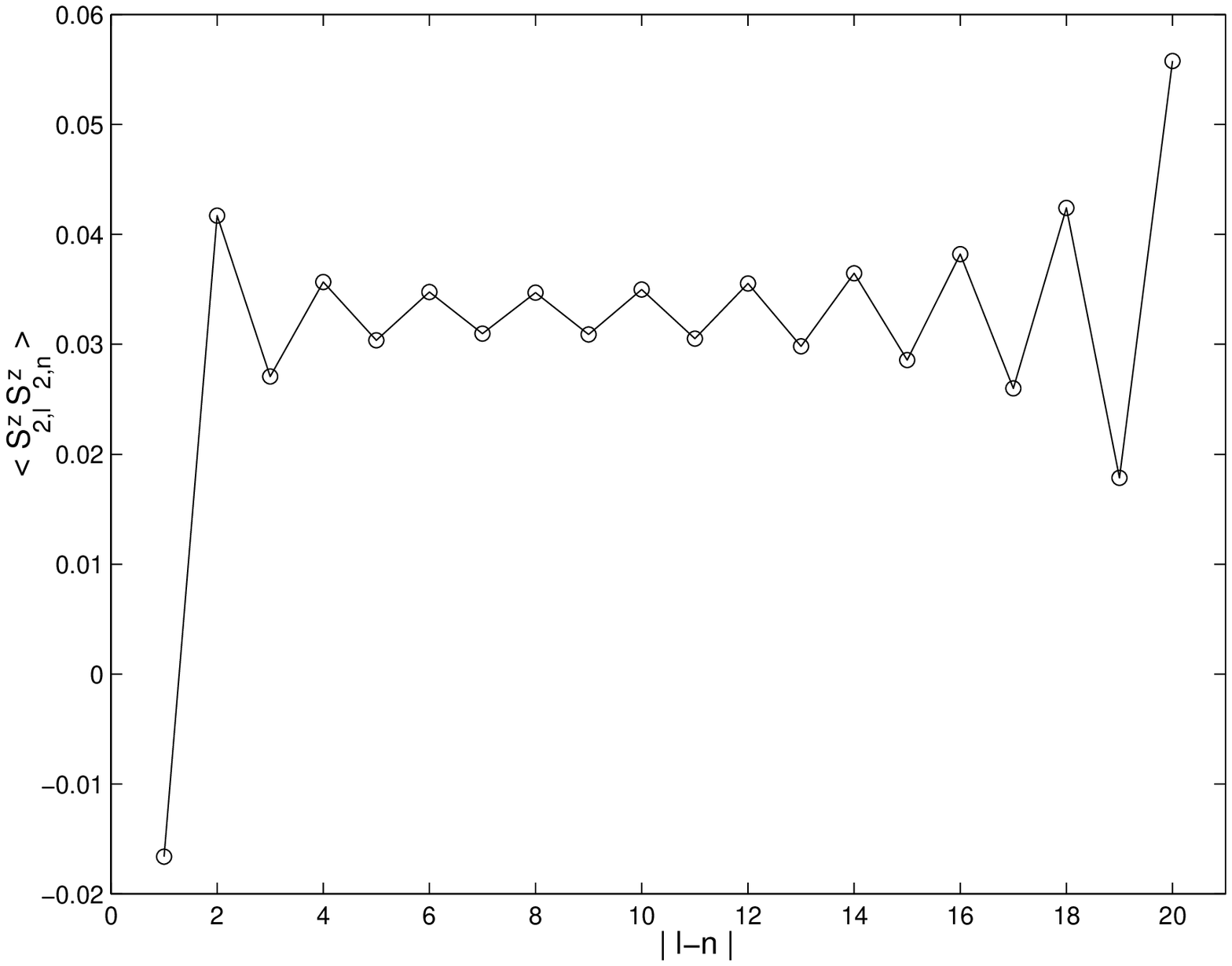,width=10cm}
\end{center}
\caption{}
\label{corr22zz}
\end{figure}

\vspace*{1cm}
\begin{figure}[hp]
\begin{center}
\epsfig{figure=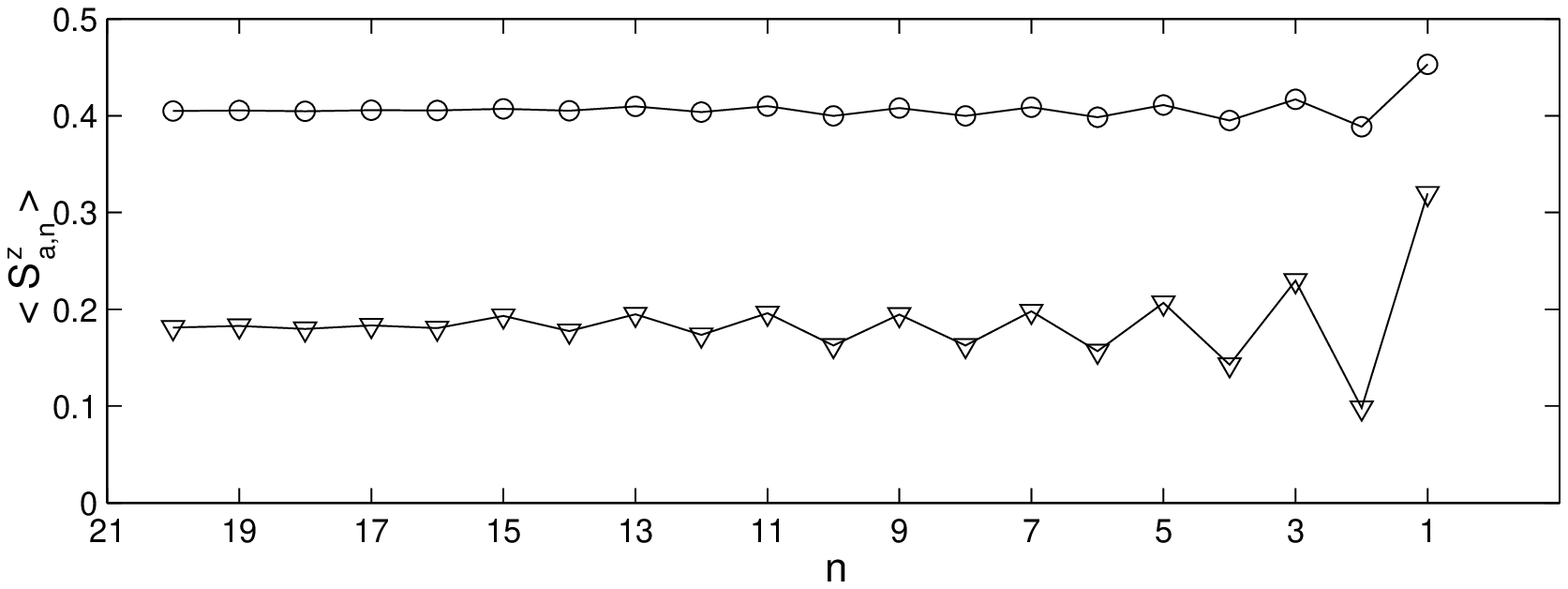,width=12cm}
\end{center}
\caption{}
\label{spnden2}
\end{figure}

\begin{figure}[hp]
\begin{center}
\epsfig{figure=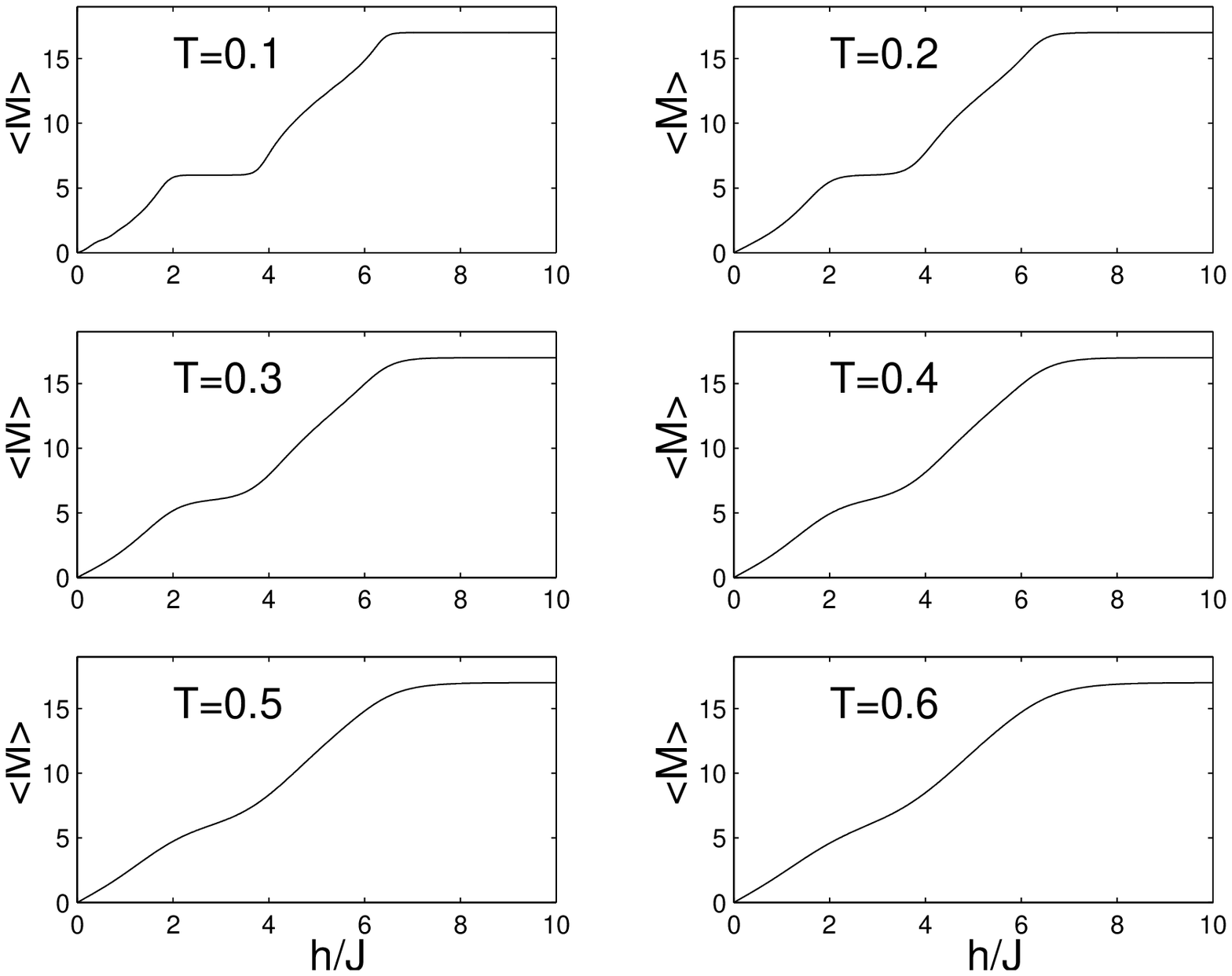,width=11cm}
\end{center}
\caption{}
\label{mo36}
\end{figure}

\begin{figure}[hp]
\begin{center}
\epsfig{figure=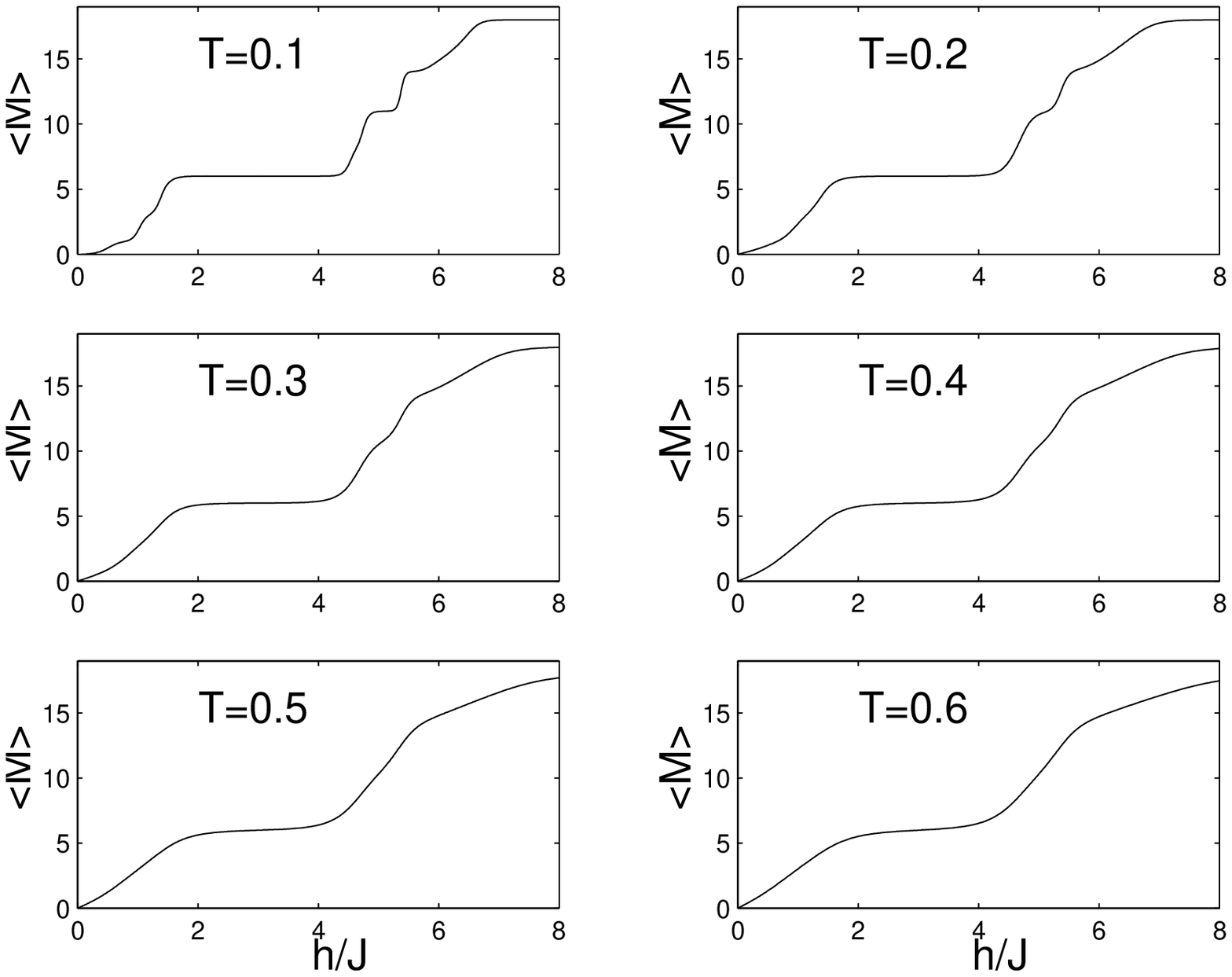,width=11cm}
\end{center}
\caption{}
\label{mp36}
\end{figure}

\begin{figure}[hp]
\begin{center}
\epsfig{figure=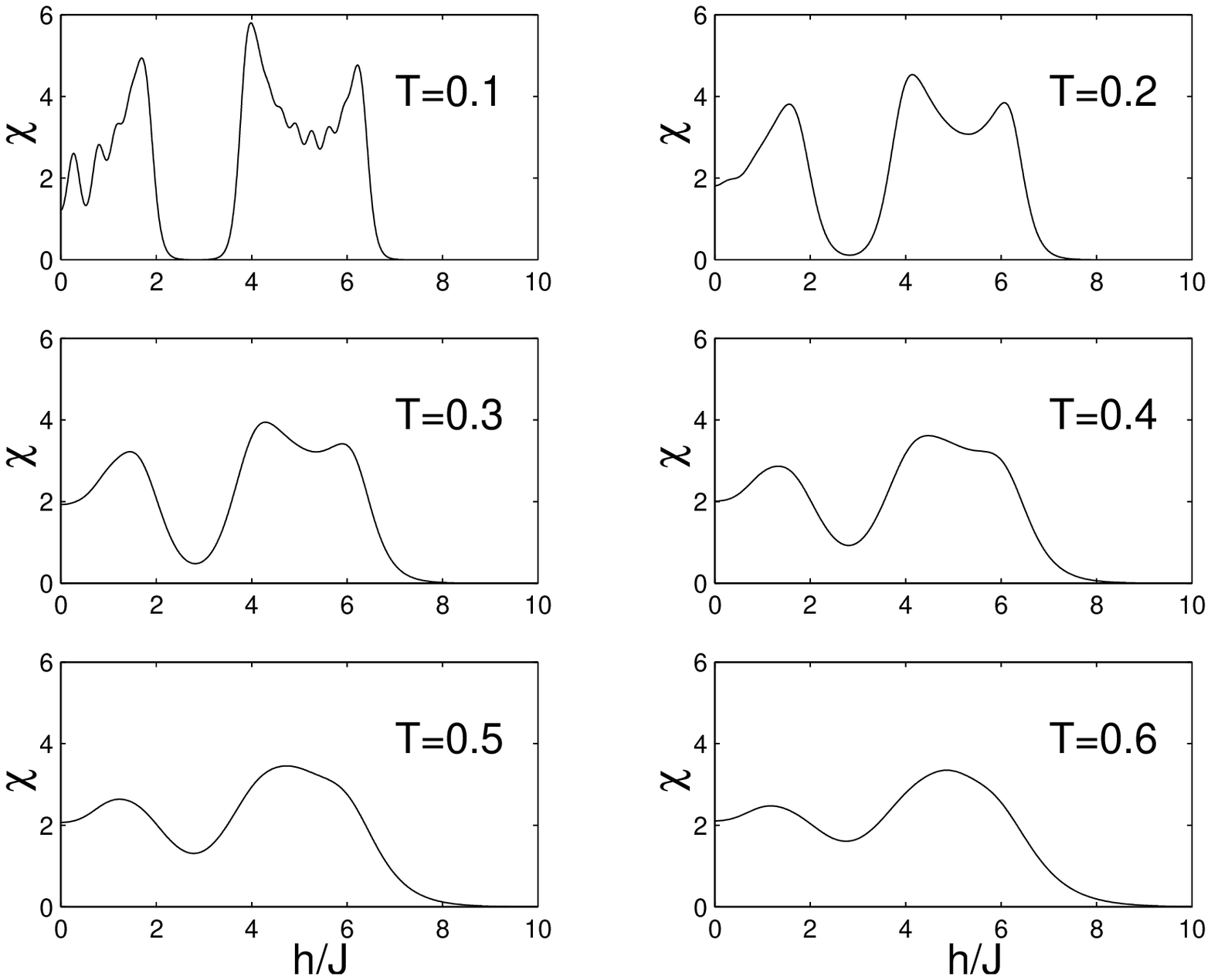,width=11cm}
\end{center}
\caption{}
\label{chio36}
\end{figure}

\begin{figure}[hp]
\begin{center}
\epsfig{figure=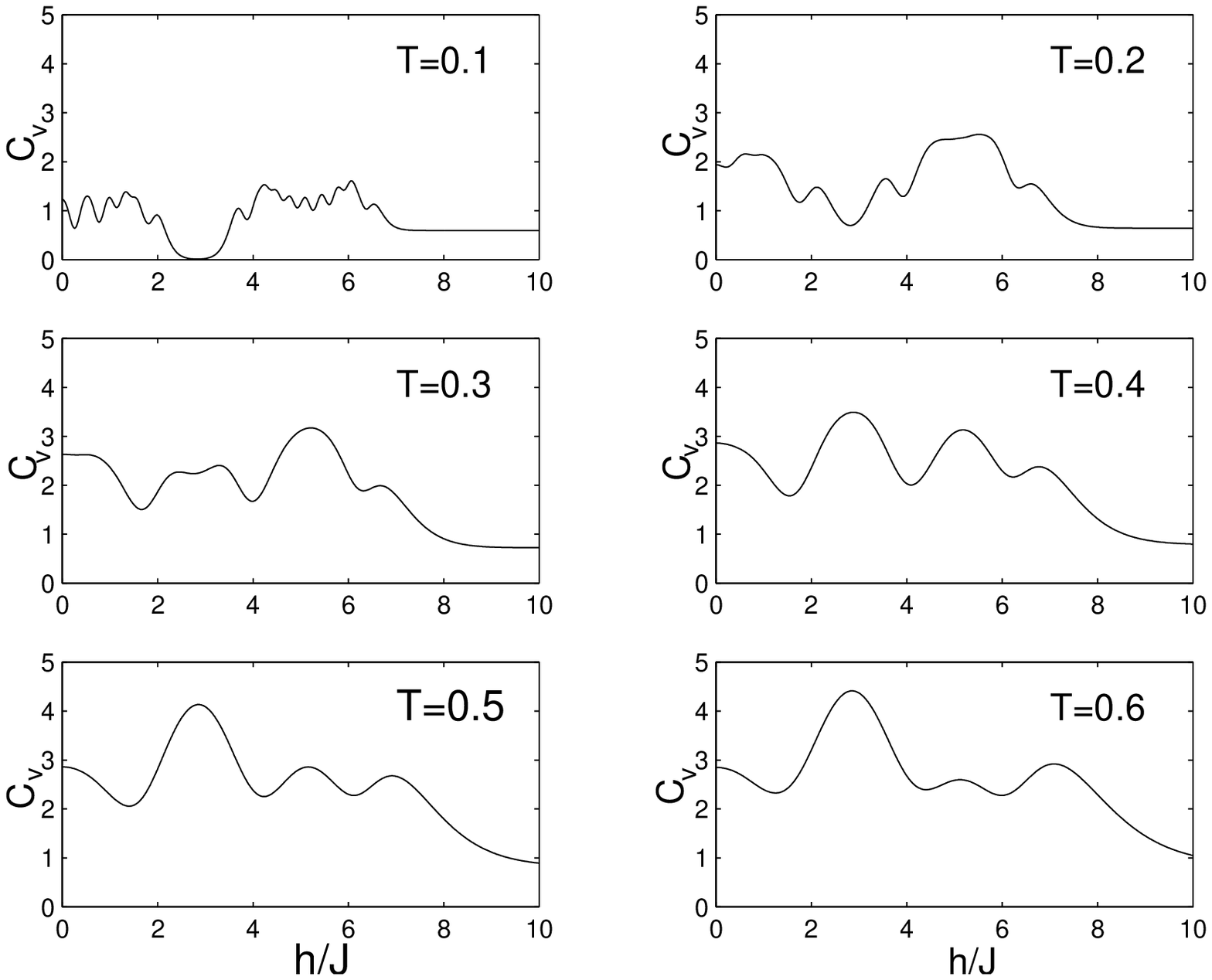,width=11cm}
\end{center}
\caption{}
\label{cvo36}
\end{figure}

\begin{figure}[hp]
\begin{center}
\epsfig{figure=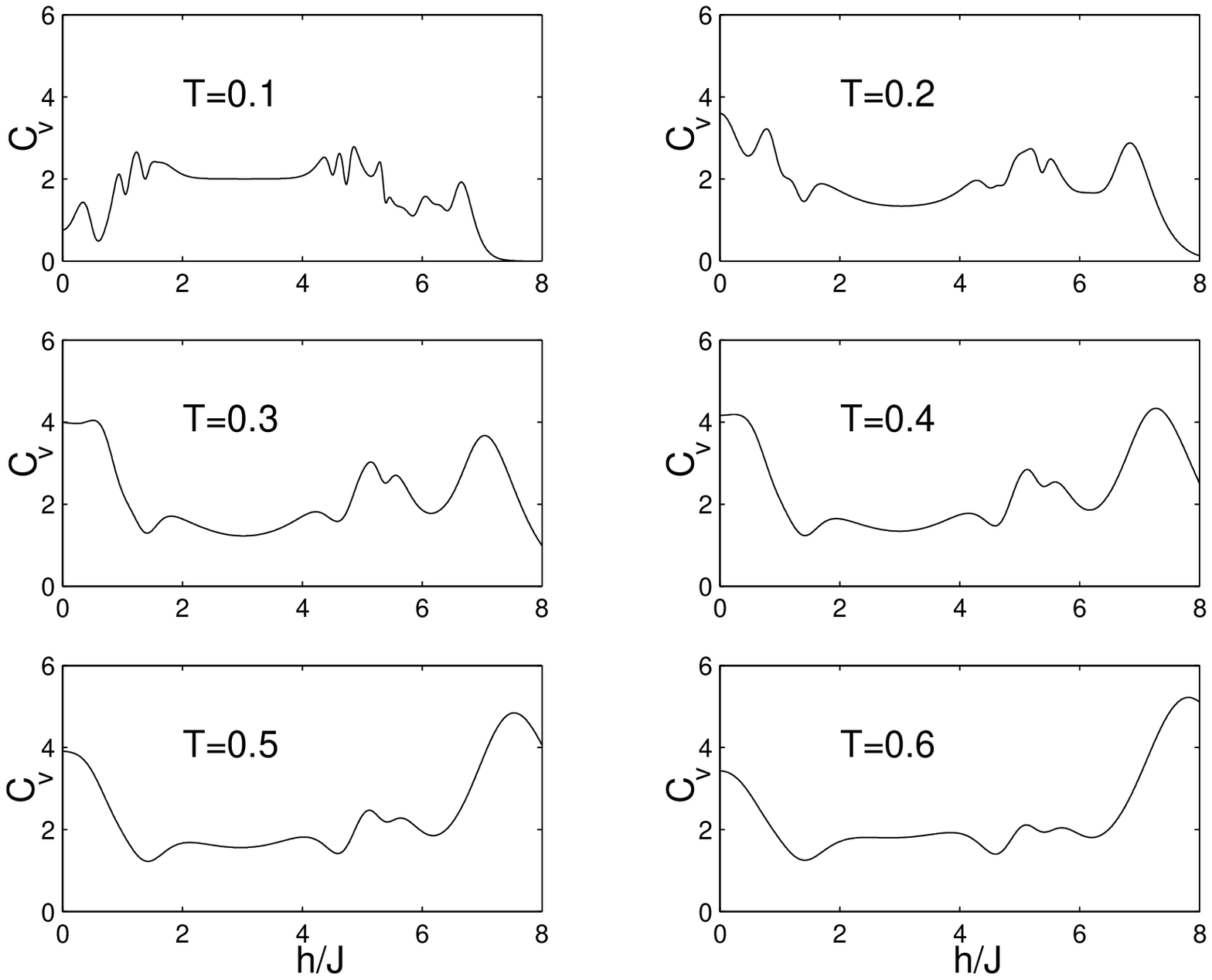,width=11cm}
\end{center}
\caption{}
\label{cvp36}
\end{figure}

\begin{figure}[hp]
\begin{center}
\epsfig{figure=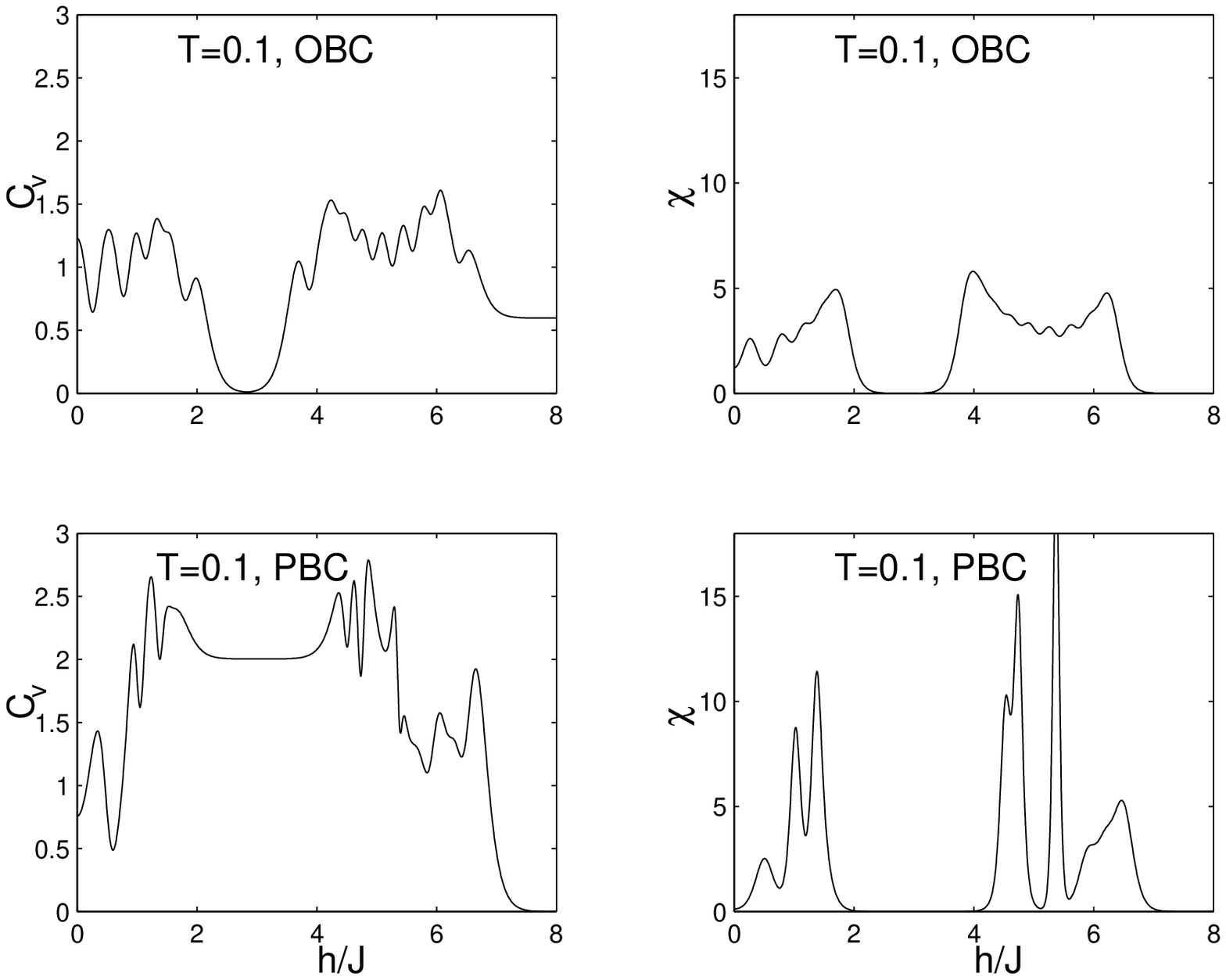,width=11cm}
\end{center}
\caption{}
\label{cvchi36}
\end{figure}

\begin{figure}[hp]
\begin{center}
\epsfig{figure=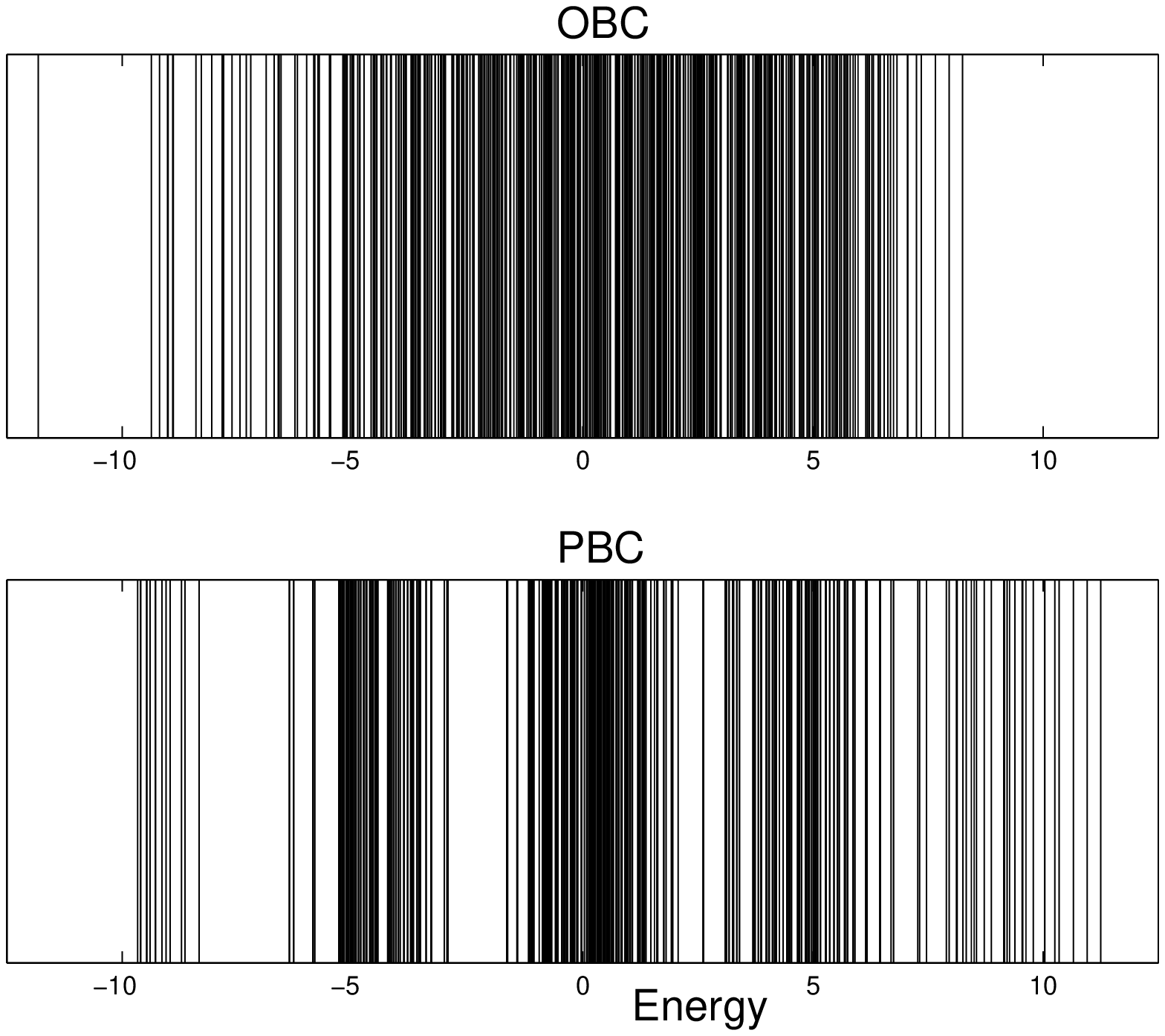,width=12cm}
\end{center}
\caption{}
\label{nrg12}
\end{figure}

\begin{figure}[ht]
\begin{center}
\epsfig{figure=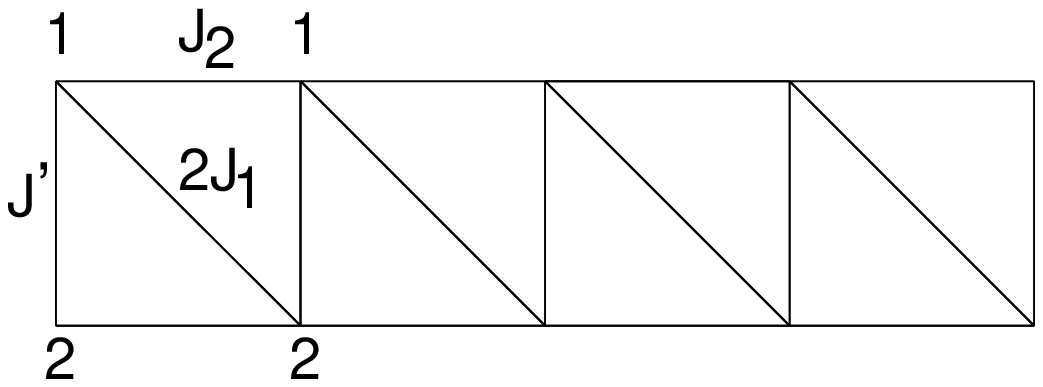,height=8cm,width=12cm}
\end{center}
\caption{}
\label{ladd2}
\end{figure}

\begin{figure}[hp]
\begin{center}
\epsfig{figure=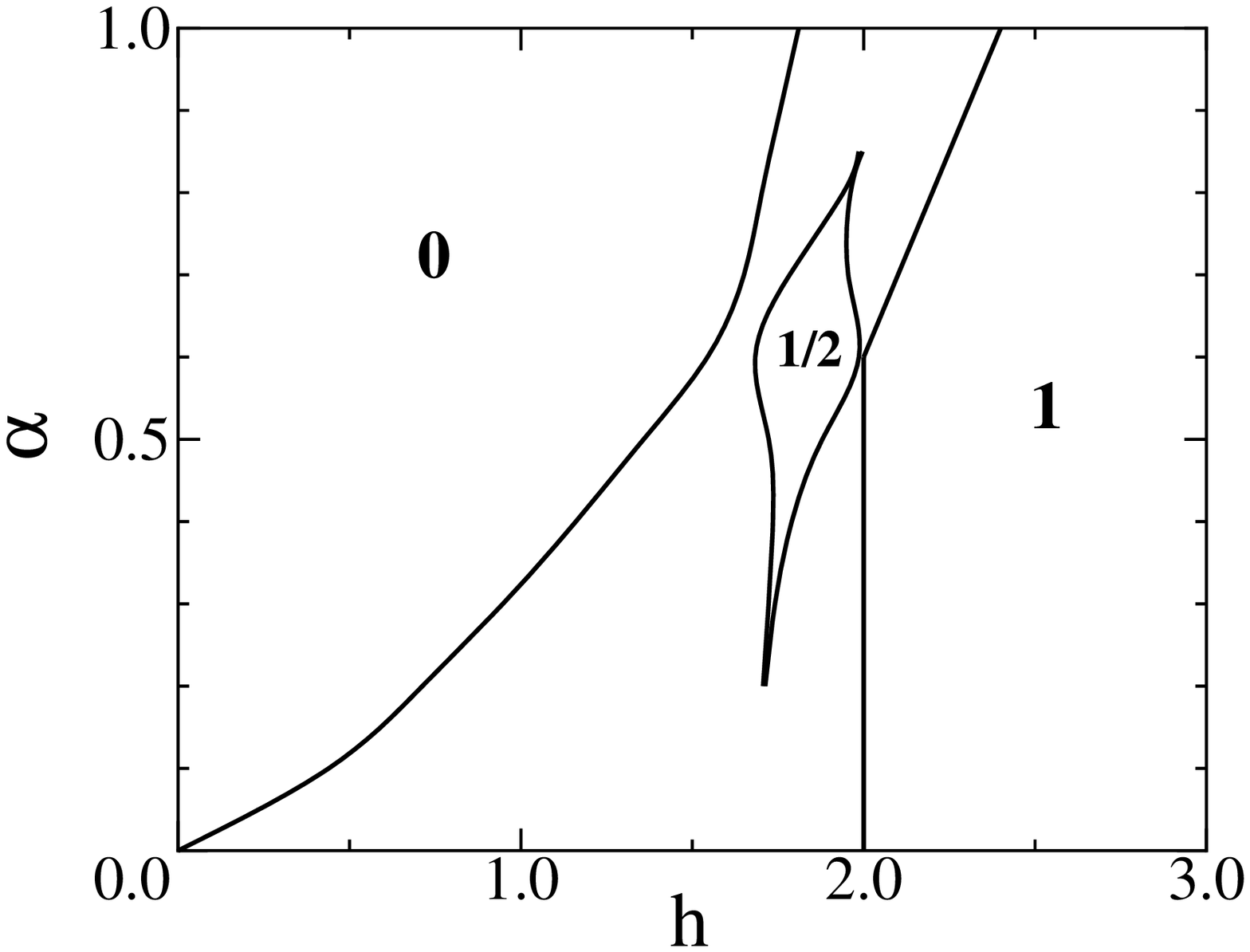,width=10cm}
\end{center}
\caption{}
\vspace*{-6cm}
\label{tone}
\end{figure}

\end{document}